\documentclass[a4paper]{article}
\usepackage{amsmath}    
\usepackage{graphicx}   
\usepackage{amssymb,amscd}    
\usepackage{enumitem}
\usepackage{color}
\usepackage{mathrsfs}
\usepackage{bm}
\usepackage{bbm,wasysym}
\usepackage{multirow}
\usepackage{easyReview}
\usepackage{array}
\usepackage{empheq}
\usepackage{pifont}
\usepackage{soul}
\usepackage{svg}
\usepackage{float}

\usepackage{jheppub} 
\usepackage{lineno} %\linenumbers

\arxivnumber{2511.05649}

\newcommand{\fc}[1]{\mathcal{#1}}

\newcommand{\mc}[1]{\text{\fontencoding{U}\fontfamily{boondoxuprscr}\fontshape{n}\selectfont #1}}
\newcommand{\mf}[1]{\mathfrak{#1}}

\newcommand{\sym}{\text{Sym}}

\renewcommand{\l}{\mathcal{L}}
\newcommand{\s}{\mathcal{S}}

\newcommand{\h}{\mathcal{H}}

\renewcommand{\d}{\partial}
\newcommand{\8}{\infty}
\newcommand{\<}{\langle}
\renewcommand{\>}{\rangle}

\newcommand{\p}{\hspace*{5ex}}
\newcommand{\ps}{\hspace*{2ex}}

\newcommand{\OO}{\fc{O}}
\newcommand{\twid}[1]{\widetilde{#1}}
\newcommand{\eps}{\varepsilon}

\newcommand{\cmt}[1]{}
\newcommand{\cl}{\text{cl}}
\newcommand{\qm}{\text{qm}}
\newcommand{\mlb}{\left\{\!\!\left\{}
\newcommand{\mrb}{\right\}\!\!\right\}}

\DeclareFontFamily{U}{boondoxuprscr}{\skewchar \font =45}
\DeclareFontShape{U}{boondoxuprscr}{m}{n}{
	<-> BOONDOXUprScr-Regular}{}
\DeclareFontShape{U}{boondoxuprscr}{b}{n}{
	<-> BOONDOXUprScr-Bold}{}

\newcommand{\bt}{\, \text{\scriptsize \textbullet}\, }
\begin{document}
%=============================================================

\title{Magnusian: Relating the Eikonal Phase, the On-Shell Action, and the Scattering Generator}

\author[a,b]{Jung-Wook Kim,}
\author[b,c]{Raj Patil,}
\author[b]{Trevor Scheopner,}
\author[b]{and Jan Steinhoff}

\affiliation[a]{Theoretical Physics Department, CERN, 1211 Geneva 23, Switzerland}
\affiliation[b]{Max Planck Institute for Gravitational Physics (Albert Einstein Institute), Am M\"{u}hlenberg 1, 14476 Potsdam, Germany}
\affiliation[c]{Institut f\"{u}r Physik und IRIS Adlershof, Humboldt Universit\"{a}t zu Berlin, Zum Gro{\ss}en Windkanal 2, 12489 Berlin, Germany}

\emailAdd{jung-wook.kim@cern.ch}
\emailAdd{raj.patil@aei.mpg.de}
\emailAdd{trevor.scheopner@aei.mpg.de}
\emailAdd{jan.steinhoff@aei.mpg.de}

\abstract{
    Two fundamentally distinct types of quantities are both called ``eikonal'' in present amplitudes literature. The unitarity of the S-matrix ensures it can be written as the exponential of a Hermitian operator. The eikonal generator or \emph{Magnusian}, which is the classical limit of the expectation value of that operator, generates all scattering observables. The leading order classical behavior of the phase of an S-matrix element is called the classical eikonal phase, and it coincides with a classical on-shell action. We demonstrate that the eikonal generator (Magnusian) and the eikonal phase (classical on-shell action) are \emph{inequivalent} and find the exact general relationship between them. That relationship explains the special case of integrable scattering in which the two do coincide up to a Legendre transformation and explains why such a correspondence fails in general when spin or radiation are included. 
}

\begin{flushright}
\begingroup\footnotesize\ttfamily
 CERN-TH-2025-226 \\
 HU-EP-25/37-RTG
\endgroup
\end{flushright}

\maketitle
\flushbottom

\section{Introduction}

    We are living in a golden age of gravitational physics. Since the first observation of gravitational waves by the LIGO/Virgo collaboration~\cite{LIGOScientific:2016aoc, LIGOScientific:2017vwq}, an increasing number of subsequent events have been detected \cite{LIGOScientific:2025slb,
LIGOScientific:2025jau,LIGOScientific:2025pvj}. The future of observations is even more exciting as promising next-generation gravitational wave detectors making progress toward realization~\cite{LISA:2017pwj,ET:2025xjr,Punturo:2010zz, Ballmer:2022uxx, Reitze:2019iox}. The burgeoning physics of gravitational wave observations carries important consequences for astronomy, cosmology, and particle physics.

    In recent years a multitude of techniques and ideas from theoretical high energy physics, in particular those of the scattering amplitudes and effective field theory community, have found significant application to the theoretical challenges posed by this new gravitational era. In the weak-field slow-velocity regime of post-Newtonian (PN) calculations, the EFT diagrammatic approach has been very useful for computing the effective conservative two-body Hamiltonian \cite{Goldberger:2004jt,Foffa:2012rn,Foffa:2019rdf,Foffa:2016rgu,Foffa:2019yfl,Blumlein:2020pog,Foffa:2019hrb,Blumlein:2021txe,Porto:2024cwd,Blumlein:2021txj,Blumlein:2020znm,Porto:2005ac,Levi:2015msa,
Kim:2021rfj,Levi:2020uwu,Levi:2019kgk,Kim:2022pou,Kim:2022bwv,Levi:2022dqm,Levi:2022rrq,Mandal:2022nty,Mandal:2022ufb} and also the dissipated energy, angular momentum flux, and the waveform \cite{Goldberger:2009qd,Ross:2012fc,
Cho:2021mqw,Cho:2022syn,Amalberti:2023ohj,Amalberti:2024jaa,Mandal:2024iug}. For the case of high-velocities in the weak-field regime, analyzed using post-Minkowskian (PM) computations, tremendous progress has been made using both scattering amplitudes~\cite{Cheung:2018wkq,Bern:2019nnu,Bern:2022kto,Bern:2024adl,Kosower:2018adc,Bjerrum-Bohr:2018xdl,Cristofoli:2019neg,Damgaard:2019lfh,Brandhuber:2021eyq,Vines:2017hyw,Bern:2021dqo,Bern:2021yeh,Damgaard:2023ttc,Bern:2023ccb,Bern:2025zno,Herderschee:2023fxh,Brandhuber:2023hhy,Georgoudis:2023lgf,Cristofoli:2021vyo} and worldlines~\cite{Mogull:2020sak,Dlapa:2021npj,Jakobsen:2021zvh,Dlapa:2021vgp,Dlapa:2023hsl,Jakobsen:2023ndj,Jakobsen:2023pvx,Kalin:2020mvi,Kalin:2020fhe,Jakobsen:2022psy,Kalin:2022hph,Jakobsen:2023hig,Dlapa:2022lmu,Dlapa:2024cje,Dlapa:2025biy,Driesse:2024xad,Driesse:2024feo,Jakobsen:2021smu,Mougiakakos:2021ckm,Bohnenblust:2023qmy,Bohnenblust:2025gir}. These developments are largely due to a blend of a clever and efficient organization of perturbative calculations and formal mathematical developments in understanding the properties of multi-loop integrals \cite{Parra-Martinez:2020dzs,Weinzierl:2022eaz,Abreu:2022mfk,Blumlein:2022qci,Frellesvig:2023bbf,Frellesvig:2024zph,Klemm:2024wtd}. Several sophisticated techniques from quantum field theory have been used to analyze the two-body problem, such as generalized unitarity \cite{Bern:1994cg,Bern:2011qt}, double copy~\cite{Bern:2010ue,Bern:2019prr,Bern:2022wqg,Adamo:2022dcm}, supersymmetry~\cite{Jakobsen:2021zvh}, massive higher spins \cite{Chung:2018kqs,Arkani-Hamed:2019ymq,Bern:2020buy,Bautista:2021wfy,Bautista:2022wjf,Chiodaroli:2021eug,Aoude:2023vdk,Cangemi:2023ysz,Akpinar:2025bkt,FebresCordero:2022jts}, Compton amplitudes \cite{Vines:2017hyw,Guevara:2019fsj,Guevara:2018wpp}, twistor variables \cite{Kim:2021rda,Kim:2024grz}, recursion relations \cite{Cho:2023kux,Damgaard:2024fqj}, etc. These techniques have made the calculation of observables insightful and systematic. 
    
In this article we focus on one of the most fundamental concepts linking quantum and classical physics: the eikonal. The eikonal was originally motivated by geometric optics and later played a crucial role in the formulation of quantum mechanics. It has also been used recently to analyze the two-body problem. But, there are two competing and incompatible usages of the name ``eikonal'' in present amplitudes literature. The first is proportional to the expectation value of the operator logarithm of the S-matrix and in the classical limit becomes a function whose derivatives contain all classical scattering observables. The second is proportional to the logarithm of an S-matrix element and in the classical limit becomes the on-shell action of the corresponding classical theory. Even in the classical limit, the two do not agree. In the simple case of an integrable system, they coincide (up to a Legendre transformation). In this article we prove the non-trivial relationship between the two quantities in general. To understand that relationship, we need to put them in the context of four closely related but conceptually distinct quantities present in the literature which have the intent of capturing the physics of perturbative classical scattering. \\

\noindent\textbf{(I) The On-Shell Action.} For a generic classical system, the action functional $\s$ evaluated on the trajectory that solves the equation of motion (on-shell) is called the \emph{on-shell action}, or Hamilton's principal function. The on-shell action is a function of the boundary conditions (initial and final coordinates) of the trajectory, and through its derivatives via the Hamilton-Jacobi conditions, it determines the entire classical dynamics. Because it depends on boundary conditions, the classical on-shell action is naturally an ``in-out'' quantity.\\

\noindent\textbf{(II) The (Optical) Eikonal Phase.} The \emph{eikonal approximation} is the starting point for geometric optics, where the wave equation is solved by an ansatz $A e^{i\delta}$ with the \emph{optical eikonal phase} $\delta$ having much larger derivatives than the wave-amplitude $A$ and small second derivatives compared to the products of its first derivatives; see e.g. \textsection 53 of Ref.~\cite{Landau:1975pou} for details. This is the traditional use of the term \emph{eikonal phase},\footnote{There are several variations in the definition of the eikonal phase in literature which are distinct outside of the classical limit and at high loop order. In the definition we use here, the impact parameter space, not necessarily defined as the plane orthogonal to incoming momenta, is used to resum $2 \to 2$ scattering amplitude which can be interpreted as an instance of the WKB approximation in quantum field theory.} as in Refs.~\cite{Amati:1990xe, Amati:2007ak, Akhoury:2013yua, KoemansCollado:2019ggb, DiVecchia:2021bdo, Brandhuber:2021eyq, Parra-Martinez:2020dzs, Laenen:2008gt, Haddad:2021znf}. For a more complete list of references, consult the review by Di Vecchia, Heissenberg, Russo, and Veneziano~\cite{DiVecchia:2023frv}. In the geometric optics limit, $\delta$ satisfies the Hamilton-Jacobi equation with an implied effective Hamiltonian for the particle-like trajectories of wave-rays. When applied to quantum mechanical scattering, in which case this is often called the WKB approximation~\cite{Sakurai:2011zz}, the implied effective Hamiltonian is the actual classical Hamiltonian of the classical limit of the quantum theory. This ``opticomechanical analogy'' led to Schr\"odinger's wave mechanics in the early developments of the quantum theory~\cite{Schrodinger:1926iou}. Thus, the optical eikonal phase $\delta$ in an appropriately taken classical limit becomes the classical on-shell action $\s$. In the context of quantum scattering, the geometric optics limit is managed by the $\hbar\to 0^+$ limit, and the wave ansatz is $A e^{\frac{i}{\hbar}\delta}$. Time evolution is performed by the S-matrix, so that the classical optical eikonal phase for propagation from an initial state $|Q_i\>$ to a final state $|Q_f\>$ is given precisely by:
\begin{equation}
    i\delta = \lim_{\hbar\to 0^+}\hbar\ln(\<Q_f|\hat{\text{S}}|Q_i\>). \label{eq:deltadef}
\end{equation}

\noindent\textbf{(III) The Eikonal Generator (Magnusian).} The \emph{eikonal generator} or \emph{Magnusian} $\chi$ is the function on classical phase space $(Q, P)$ so that the outgoing value of any observable $A(Q,P)$ in a scattering event is related to its incoming value via the exponentiated action of the Poisson bracket:
\begin{equation}
    A_\text{out} = e^{\{\chi,\bt\}} A_\text{in}.  \label{eq:clinout}
\end{equation}
This property of the eikonal generator (Magnusian) makes the operator $e^{\{\chi,\bt\}}$ the classical equivalent of the S-matrix. It has a systematic perturbative expansion in the interaction picture~\cite{Kim:2024grz, Kim:2024svw} using the Magnus expansion~\cite{Magnus:1954zz, KRUGER197536}.

The Magnusian has been called ``the eikonal'' in several papers, for example in Refs.~\cite{Bern:2020buy, Bern:2023ity, Gatica:2023iws, Luna:2023uwd, Bern:2021xze, Bohnenblust:2024hkw, delaCruz:2021gjp,Kim:2025olv,Kim:2025hpn,Bern:2022kto}. It has also been called ``the on-shell action'', see for example Refs.~\cite{Haddad:2025cmw, He:2025how}. Whenever we say eikonal generator or Magnusian, we always mean $\chi$. We will never refer to $\chi$ simply as ``the eikonal'', because that name is ambiguous and does not coincide with the quantity defined in (II). We will also never call $\chi$ the ``the on-shell action'' because that does not coincide with the quantity defined in (I). 

    To understand how the eikonal generator (Magnusian) arises classically, it is simplest to approach it from quantum mechanics. In quantum mechanics, the $\hat\chi$ operator is defined by the exponential representation of the S-matrix~\cite{Lehmann:1957zz, Damgaard:2021ipf, Damgaard:2023ttc}:
    \begin{equation}
        \hat{\text{S}} = e^{\frac{i}{\hbar}\hat \chi}. \label{eq:chifromsmatrix}
    \end{equation}
    The unitarity of the S-matrix ensures the Hermiticity of $\hat\chi$. The S-matrix performs the entire time-evolution of the scattering process so that $|\psi_{\text{in}}\>$ evolves to $|\psi_{\text{out}}\> = \hat{\text{S}}|\psi_{\text{in}}\>$. The outgoing expectation value $\<\psi_\text{out}|\hat A|\psi_{\text{out}}\>$ for any operator $\hat A$ is related to the incoming state $|\psi_\text{in}\>$ by:
    \begin{equation}
        \<\psi_\text{out}|\hat A|\psi_{\text{out}}\> = \<\psi_\text{in}|\hat{\text{S}}^\dagger\hat A\, \hat{\text{S}}|\psi_\text{in}\> = \<\psi_\text{in}|\left(e^{\frac{1}{i\hbar}[\hat\chi,\bt]} \hat A\right)|\psi_\text{in}\> \label{eq:qminout}
    \end{equation}
    where the operator exponential is defined by series expansion and $[\hat\chi,\bt] (\hat A) = [\hat\chi,\hat A]$, with $[\hat\chi,\bt]^n$ understood to mean its $n$-fold composition. While there is subtlety in taking the classical limit of S-matrix elements, the classical function $\chi$ defined by:
    \begin{equation}
        i\chi = i\lim_{\hbar\to 0^+}\<\hat\chi\> = \lim_{\hbar\to 0^+}\hbar\<\psi_\text{in}|\ln\hat{\text{S}}|\psi_\text{in}\> \label{eq:chidef}
    \end{equation}
    and the classical limit of \eqref{eq:qminout} are well-defined, where the latter yields \eqref{eq:clinout}. Although originally motivated in the context of conservative dynamics, the classical eikonal generator (Magnusian) can be generalized to incorporate radiative effects by enlarging the classical phase space to include radiation field degrees of freedom~\cite{Kim:2025hpn}. Because $\chi$ is defined by an expectation value in the initial state, it is naturally an ``in-in'' quantity.
    
    Carefully note the difference in placement of the logarithm between \eqref{eq:deltadef} and \eqref{eq:chidef}. 
    The na\"ive expectation that the two might coincide (have the same real part) in the classical limit is \emph{wrong}.
    Given $\hat\chi$, its matrix elements can be given an $\hbar$ expansion. Expanding and computing the exponential as a series in \eqref{eq:chifromsmatrix} in terms of these matrix elements gives an infinite tower of ``$\hbar$ over $\hbar$'' cancellations so that the leading classical behavior of the matrix elements of the S-matrix, which determine $\delta$, depend in a complicated way on $\chi$. 
    The inequivalence of $\chi$ and $\delta$ (under various different names) has been noted before~\cite{Kalin:2020mvi, Damgaard:2021ipf, Bern:2021dqo, Haddad:2025cmw}. We, for the first time, derive the exact relationship between them in general. Our main result will be to prove that in the classical limit:
    \begin{equation}
        \delta = \s(Q_f, Q_i) = \frac{e^{\{\chi,\bt\}}-1}{\{\chi,\bt\}}(P_i\{\chi, Q_i\}) + \chi(Q_i, P_i). \label{eq:main}
    \end{equation}

\noindent\textbf{(IV) The Radial Action.} In the case of an integrable classical scattering system, such as the conservative sector of the PM or PN gravitational scattering of two nonspinning bodies, the Hamilton-Jacobi equation is separable and observables can be computed from derivatives of the radial component $I$ of the on-shell action. This has been observed previously, by for example Refs.~\cite{Kalin:2020mvi, Bern:2021dqo, Gonzo:2024zxo}. The authors of~\cite{Gonzo:2024zxo} showed in the case of a spinless probe in a Kerr background (which is integrable), that $I$ can be used in equation \eqref{eq:clinout} in place of $\chi$. From the exact relationship between $\chi$ and $\s$ [eq.~\eqref{eq:main}], we show in sec.~\ref{sec:intsys} that for general integrable systems the two are related by a simple Legendre transform, and consequently for integrable scattering $\chi$ and $I$ coincide. \\

An exactly-solvable example where $\s$ and $\chi$ are distinct is the simple harmonic oscillator, where we treat the kinetic energy term as the (solvable) ``free'' Hamiltonian $\h_0$ and the potential energy term as the perturbation term (interaction Hamiltonian) $\h_{I}$,
\begin{align}
    \h = \frac{p^2}{2} + \lambda \frac{q^2}{2} = \h_0 + \lambda \h_{I} \,.
\end{align}
As we explicitly compute in sec.~\ref{sec:SHOex}, the on-shell action and the eikonal generator computed using the classical interaction picture~\cite{Kim:2025olv} yield \emph{different} results for the time evolution from $t = 0$ to $t = T$: the on-shell action $\s$ (expressed as a function of initial conditions $Q_i$ and $P_i$) and the eikonal generator computed using the Magnus series~\cite{Kim:2024svw} are:
\begin{align}
    \s (T,0,Q_i, P_i) &= \frac{1}{6} \lambda  \left(P_i^2 T^3-3 Q_i^2 T\right) -\frac{1}{30} \lambda ^2 T^3 \left(3 P_i^2 T^2+10 P_i Q_i T+5
   Q_i^2\right) \nonumber
   \\ &\phantom{=asdfasdfasdf} +\frac{1}{630} \lambda ^3 T^5 \left(10
   P_i^2 T^2+56 P_i Q_i T+63 Q_i^2\right) + \mathcal{O} (\lambda^4) \,.\\
    \chi (T,0,Q_i,P_i) &= -\frac{1}{6} \lambda  T \left(P_i^2 T^2+3 P_i Q_i T+3 Q_i^2\right) +\frac{1}{60} \lambda ^2 T^3 \left(P_i^2 T^2+5 P_i Q_i T+5 Q_i^2\right)\nonumber
   \\ &\phantom{=asdfasdfasdfasdf}-\frac{1}{3780}\lambda ^3 T^5 \left(11
   P_i^2 T^2+42 P_i Q_i T+42 Q_i^2\right) + \mathcal{O} (\lambda^4) \,,
\end{align}
which are clearly not the same. However, they are not independent and are related precisely by \eqref{eq:main}. 

This paper is organized as follows. First in sec. \ref{sec:review} we review the basics of the classical definitions of the on-shell action and the eikonal generator, which going forward for brevity we call the Magnusian, and give an introduction to the classical interaction picture. Sec.~\ref{sec:result} contains the main results of this paper; here we provide both a completely classical and a completely quantum proof of how the on-shell action and the Magnusian are related and recover the well-known result that the classical limit of the optical eikonal phase is the on-shell action. In sec.~\ref{sec:properties}, we discuss the interpretation of the Magnusian, study its symmetries, and provide perturbative simplifications of the exact relations between it and on-shell action. In sec.~\ref{sec:exandapp} we apply our results to a few examples, which give simple illustrations that $\chi\neq \s$ and which show why $\chi = I$ in the case of integrable scattering. Finally, we present our conclusions and avenues for future work in sec.~\ref{sec:conclusion}. Appendix~\ref{sec:wigner} details the classical limit of Wigner functions and appendix~\ref{sec:quant} details the classical limit of the path integral. A further example of motion in $SU(2)$ which helps illustrate the relationship between the Magnusian and the Hamiltonian is given in appendix~\ref{sec:su2}. With this article we also attach a Mathematica file \texttt{harmonic\_oscillator.m} where the relationship between the Magnusian and on-shell action is verified for the oscillator cases to higher orders. 

\section{Review of the On-Shell Action and the Eikonal Generator}\label{sec:review}

To set up notation, we review the on-shell action, the Magnusian~\cite{Kim:2024grz}, and the Magnus series~\cite{Magnus:1954zz} which is used to compute the Magnusian in the classical interaction picture~\cite{Kim:2024svw,Kim:2025olv}. 

\subsection{The On-Shell Action}

For a classical system of coordinates $Q(t)$, canonically conjugate momenta $P(t)$, and Hamiltonian $\h(t, Q, P)$, the action is fundamentally defined as the functional:
\begin{equation}
    \s[Q, P](t_f, Q_f, t_i, Q_i) = \int_{t_i}^{t_f}(P\dot Q - \h(t, Q, P)) dt
\end{equation}
where the bracket versus parentheses notation $[], ()$ indicates that the action is a functional of the phase space path $(Q(t), P(t))$ while it is an ordinary function of the boundary times $t_i, t_f$ and boundary coordinates $Q_i = Q(t_i)$ and $Q_f = Q(t_f)$. When considering symmetries of the action, it will also be useful to consider it as a functional of the Hamiltonian. Under a variation of all its dependencies, the action satisfies:
\begin{equation}
\delta \s = P_f \delta Q_f - P_i \delta Q_i - \h_f \delta t_f + \h_i \delta t_i +\! \int_{t_i}^{t_f}\!\!\left(\delta P\left(\dot Q - \frac{\d\h}{\d P}\right) - \delta Q\left(\dot P + \frac{\d\h}{\d Q}\right) - \Delta\h\right) dt \label{eq:actvariation}
\end{equation}
where $\Delta\h$ represents a possible variation in the \emph{functional form of the Hamiltonian}, not simply a change in its \emph{value}. When the path $(Q(t), P(t))$ is placed on-shell, we have Hamilton's equations:
\begin{equation}
    \frac{dQ}{dt} = \{Q, \h\} = \frac{\d \h}{\d P}, \p \frac{dP}{dt} = \{P, \h\} = -\frac{\d\h}{\d Q}, \label{eq:hamilton}
\end{equation}
and when the the action is evaluated on the on-shell path, the resulting function, known as the on-shell action (OSA) or Hamilton's principal function, thus has the exact differential under changes in the boundary conditions:
\begin{equation}
    d\s = P_f dQ_f - P_i dQ_i - \h_f dt_f + \h_i dt_i \label{eq:hamjacdiff}
\end{equation}
where $\h_f = \h(t_f, Q_f, P_f)$ and $\h_i = \h(t_i, Q_i, P_i)$. This exact differential is identical to the Hamilton-Jacobi conditions:
\begin{equation}
    P_f = \frac{\d \s}{\d Q_f}, \qquad P_i = -\frac{\d \s}{\d Q_i}, \qquad \h_f = -\frac{\d \s}{\d t_f}, \qquad \h_i = \frac{\d \s}{\d t_i}. \label{eq:hamjac}
\end{equation}
Formally, given solutions to the classical equations of motion, integrating these conditions determines the OSA and in converse, knowing the OSA determines the solutions to the classical equations of motion through these conditions. The Hamilton-Jacobi conditions are exactly equivalent to the statement that the classical OSA is the generating function of the canonical transformation corresponding to time evolution from $t_i$ to $t_f$. 

The OSA is properly a function of the \emph{boundary conditions} $(Q_i, Q_f)$ as independent variables. However, for comparing with the Magnusian (in examples, to directly see their inequivalence), it will be convenient to use the Hamilton-Jacobi conditions to express it in terms of the \emph{initial conditions} $(Q_i, P_i)$. Importantly, the on-shell action expressed as a function of initial conditions $\s(t_f, t_i, Q_i, P_i)$ does not act properly as a generating function in any sense. 

\subsection{The Eikonal Generator (Magnusian)}

Whatever the actual Hamiltonian $\h(t, Q, P)$ is, for fixed $t_i$ and $t_f$, the relationship between $(Q_i, P_i)$ and $(Q_f, P_f)$ is a canonical transformation. In particular, it is a canonical transformation in the subgroup of canonical transformations which are continuously connected to the identity. The group of such canonical transformations is a connected Lie group and therefore there exists an element of the Lie algebra which performs that canonical transformation under exponentiation.\footnote{Technically, this requires the canonical transformation be within a neighborhood of the identity so that the logarithm in the Lie group is well-defined. This is precisely the same condition necessary for the on-shell action to determine the canonical transformation through the Hamilton-Jacobi conditions, 
which is the context we are interested in.
} The elements of the Lie algebra of canonical transformations are Hamiltonian vector fields, i.e. differential operators of the form $\{f, \bt\}$ for a well-behaved function $f$ on phase space. So, there must exist a function $\chi$ so that the Lie algebra element $\{\chi,\bt\}$ when exponentiated gives the canonical transformation of time evolution. This is the completely classical definition of the Magnusian: 
\begin{align}
\begin{aligned}
    A(Q_f, P_f) &= \left.e^{\{\chi(Q,P),\bt\}}A(Q, P)\right|_{(Q,P) = (Q_i,P_i)} \\
    &= A(Q_i, P_i) + \{\chi, A\} + \frac{1}{2}\{\chi,\{\chi, A\}\} + \frac{1}{6}\{\chi,\{\chi,\{\chi,A\}\}\} + \cdots
\end{aligned} \label{eq:eikexpdef}
\end{align}
for $A(Q,P)$ any analytic function on phase space. Because the canonical transformation between initial and final phase space coordinates depends on $t_i, t_f$, the Magnusian is properly written as a function as $\chi(t_f, t_i, Q, P)$. The exponential representation in \eqref{eq:eikexpdef} is formally exact, while the series expression is essential for perturbation theory. 

The original method for calculating the Magnusian, the Magnus formula~\cite{Magnus:1954zz}, approaches it via the Hamiltonian. Let $\Omega(t_f, t_i)$ be the vector field on phase space associated to the Magnusian:
\begin{equation}
    \Omega(t_f, t_i) = \{\chi(t_f, t_i, Q, P), \bt\}.
\end{equation}
With $\Omega$ understood as a differential operator, \eqref{eq:eikexpdef} is equivalently stated as:
\begin{equation}
    A(Q_f, P_f) = \left. e^{\Omega(t_f, t_i)} A\right|_{(Q_i, P_i)}. 
\end{equation}
The total $t_f$ derivative of this expression (meaning the point $(Q_f, P_f)$ is also time-evolved at fixed initial conditions), combined with Hamilton's equations, gives:
\begin{equation}
    \frac{dA_f}{dt_f} = \frac{d}{dt_f}\left(e^\Omega\right) e^{-\Omega} A_f = \{A_f, \h_f\}
\end{equation}
and so we can identify that as vector fields:
\begin{equation}
    \frac{d}{dt_f}\left(e^\Omega\right) e^{-\Omega} = \{-\h_f, \bt\}.
\end{equation}
For $\eps$ a real constant parameter, it is quick to show that:
\begin{equation}
    \frac{\d}{\d\eps}\left(\frac{d}{dt_f}\left(e^{\eps\Omega}\right) e^{-{\eps\Omega}}\right) = e^{\eps\Omega}\frac{d\Omega}{dt_f} e^{-\eps\Omega} = e^{\eps[\Omega,\bt]}\left(\frac{d\Omega}{dt_f}\right).
\end{equation}
Integrating this from $\eps = 0$ to $\eps = 1$ gives immediately that:
\begin{equation}
    \{-\h_f, \bt\} = \frac{e^{[\Omega,\bt]}-1}{[\Omega,\bt]} \frac{d\Omega}{dt_f}
\end{equation}
where the operator function expression is defined precisely through Taylor series. Formally inverting this operator, reexpressing $\Omega$ in terms of $\chi$, and using the identity $[\{ A , \bt \} , \{ B , \bt\}] = \{ \{ A , B \} , \bt \}$ gives:
\begin{equation}
    \frac{d\chi(t_f, Q(t_f), P(t_f))}{dt_f} = -\frac{\{\chi,\bt\}}{e^{\{\chi, \bt\}}-1}\h_f(t_f, Q(t_f), P(t_f)). \label{eq:MagnusDE2}
\end{equation}
We are actually interested in the explicit $t_f$ dependence of the Magnusian, and so we change to the partial $t_f$ derivative by using the chain rule and Hamilton's equations (meaning the point $(Q, P)$ is no longer time evolved):
\begin{equation}
    \frac{\d \chi}{\d t_f} = \frac{d\chi}{dt_f} - \{\chi, \h_f\}
\end{equation}
which upon simplification gives:
\begin{equation}
    \frac{\d\chi(t_f, t_i, Q, P)}{\d t_f} = -\frac{\{\bt, \chi\}}{e^{\{\bt, \chi\}}-1} \h(t_f, Q, P). \label{eq:magnusformula}
\end{equation}
Note the sign difference in the nested brackets between \eqref{eq:MagnusDE2} and \eqref{eq:magnusformula}.
Repeating the same analysis for the initial time dependence gives:
\begin{equation}
    \frac{\d\chi(t_f, t_i, Q, P)}{\d t_i} = \frac{\{\chi, \bt\}}{e^{\{\chi,\bt\}}-1} \h(t_i, Q, P). \label{eq:magnusinitial}
\end{equation}
This formula can also be guessed by appreciating that interchanging the initial and final times then performs the inverse of time evolution as a canonical transformation, which by \eqref{eq:eikexpdef} must be generated by $-\chi$.

Equation \eqref{eq:magnusformula} is the \emph{Magnus formula}; it is formally exact, but is only useful for calculations when treated by series expansion. The corresponding series is the \emph{Magnus series}. The relevant expansion is:
\begin{equation}
    \frac{x}{e^x - 1} = \sum_{k=0}^\infty \frac{B_k x^k}{k!}
\end{equation}
where $B_k$ are the Bernoulli numbers (with $B_1 = - \frac{1}{2}$). With this expansion, the Magnus formula \eqref{eq:magnusformula} more properly becomes:
\begin{equation}
    \frac{\d \chi(t_f, t_i, Q, P))}{\d t_f} = \sum_{k=0}^\8 \frac{B_k}{k!}\{\bt, \chi\}^k(-\h(t_f, Q, P)) \label{eq:magnusseries}
\end{equation}
to be solved with the initial condition:
\begin{equation}
    \chi(t_i, t_i, Q, P) = 0 \label{eq:chiinit}
\end{equation}
so that $e^{\{\chi,\bt\}}$ becomes the identity transformation when no time evolution is applied. 

In practical implementations such as numerical integrators, the series \eqref{eq:magnusseries} is truncated to some finite order $m$, and the Magnusian corresponding to the finite evolution time $\Delta t = t_f - t_i$ computed from the truncated series to the $m$-th order has an error bound of $\OO (\Delta t^{m+2})$~\cite{Blanes:2008xlr}.

\subsection{The Classical Interaction Picture}

While we will prove the relationship between $\chi$ and $\s$ exactly, for most calculations of any use perturbation theory is necessary. Just as in quantum field theory, the interaction picture (Dirac picture) is the natural setting for perturbation theory, and in particular it is useful for calculating the Magnusian~\cite{Kim:2025olv}. The key idea is to separate the Hamiltonian $\h$ into a (solvable) unperturbed Hamiltonian $\h_0$ and an interaction term $\h_1$, and to describe the evolution under the full Hamiltonian $\h = \h_0 + \h_1$ by perturbative corrections to the (exact) solutions of the dynamics of the unperturbed Hamiltonian. To facilitate the perturbative expansion, it is convenient to introduce a formal expansion parameter $\lambda$ and write the full Hamiltonian in terms of phase space variables $(q, p)$ as
\begin{equation}
    \h(t, q, p) = \h_0(t, q, p) + \lambda \h_1(t, q, p)\, .
\end{equation}

Because we assume that the dynamics produced by $\h_0$ alone are explicitly solvable, we are able to compute the on-shell action $\s_0(t_f, q_f, t_i, q_i)$ corresponding to time evolution under $\h_0$ alone. In the full system, we then use $-\s_0(t, q, t_i, Q)$ as a generating function for a canonical transformation to interaction picture variables $(Q, P)$:
\begin{equation}
    P = -\frac{\d \s_0(t, q, t_i, Q)}{\d Q}, \p p = \frac{\d \s_0(t, q, t_i, Q)}{\d q}.
\end{equation}
By the Hamilton-Jacobi conditions \eqref{eq:hamjac}, if the free system was started with initial condition $Q$ then to reach final condition $q$, it would need initial momentum $P$. The new variables $(Q, P)$ time evolve using Hamilton's equations with the new Hamiltonian:
\begin{equation}
    \h'(t, Q, P) = \h(t, q, p) + \frac{\d \s_0(t, q, t_i, Q)}{\d t} = 
    \lambda\h_1(t, q(t, Q, P), p(t, Q, P)) = \lambda\h_I(t, Q, P)
\end{equation}
due to the Hamilton-Jacobi conditions. In the above, the final step defines the interaction Hamiltonian $\h_I$. With this new Hamiltonian, Hamilton's equations become:
\begin{equation}\label{eq:eominintpic}
    \frac{dQ}{dt} = \{Q, \h'\} = \lambda\frac{\d\h_I}{\d P}, \p \frac{dP}{dt} = \{P, \h'\} = -\lambda\frac{\d\h_I}{\d Q}. 
\end{equation}

Time derivatives in the interaction picture are by construction of order $\lambda$ and so give natural formal expansions in powers of $\lambda$, making the Magnus series \eqref{eq:magnusseries} become perturbatively useful. Explicitly:
\begin{align}
    \frac{\d \chi(t_f, t_i, Q, P))}{\d t_f} = -\lambda\sum_{k=0}^\8 \frac{B_k}{k!}\{\bt, \chi\}^k(\h_I(t_f, Q, P)) \,. \label{eq:magnusperturb}
\end{align}
$\chi$ is of $\OO(\lambda)$ and so each higher term in this series is suppressed by an additional order of $\lambda$. Due to this, equation \eqref{eq:magnusperturb} can be iteratively integrated to obtain the Magnusian $\chi$ as a formal explicit series in $\lambda$. 

If the interaction Hamiltonian is the interaction term of some scattering process, then because $\h_I$ will go to 0 sufficiently quickly for early and late times, the $t_f\to\8, t_i\to-\8$ limits of $\chi$ exist. Taking such limits gives that in that case, by \eqref{eq:eikexpdef}, incoming and outgoing observables exactly satisfy \eqref{eq:clinout}. It is \eqref{eq:clinout} which defines precisely what we mean when we say that the Magnusian generates all classical scattering observables. 

\section{The Action from the Magnusian} \label{sec:result}

\subsection{Classically} \label{sec:classical}

We will now prove the right half of \eqref{eq:main} which relates the on-shell action to the Magnusian completely classically. Momentarily consider a hypothetical system with time independent Hamiltonian $h(Q, P)$. Call the time parameter for this hypothetical system $\eps$. For $A(Q,P)$ any analytic function on phase space:
\begin{gather}
    \frac{dA}{d\eps} = \{A, h\} \implies \frac{d^n A}{d\eps^n} = \{-h,\bt\}^n A \\
    \implies A(Q_f, P_f) = \left.e^{\{-(\eps_f-\eps_i)h,\bt\}}A\right|_{(Q,P) = (Q_i, P_i)}\label{eq:hamexpdef}
\end{gather}
by Taylor series. The first implication is only true because we assumed $h$ is $\eps$ independent. 

Now let us return to our real system with Hamiltonian $\h(t, Q, P)$ giving true $t$ time evolution. Notice that equation \eqref{eq:hamexpdef} is nearly identical to \eqref{eq:eikexpdef}. In particular, if we consider the path $(q(\eps), p(\eps))$ through phase space satisfying:
\begin{equation}\label{eq:qpfrommagnusian}
    q(0) = Q_i, \p p(0) = P_i, \p \frac{dq}{d\eps} = -\frac{\d\chi(t_f, t_i, q, p)}{\d p}, \p \frac{dp}{d\eps} = \frac{\d\chi(t_f, t_i, q, p)}{\d q}
\end{equation}
then \eqref{eq:hamexpdef} becomes \eqref{eq:eikexpdef} where we identify $-\chi$ as the effective ``$\eps$-time''-independent Hamiltonian and the $\eps$-time evolution interval as $[0,1]$. By comparing \eqref{eq:hamexpdef} and \eqref{eq:eikexpdef}, it is automatic that:
\begin{equation}
    q(1) = Q_f, \p p(1) = P_f. 
\end{equation}
Therefore, we can equivalently define the path $(q(\eps), p(\eps))$ as satisfying the $\chi$ equations of motion with \emph{boundary conditions}:
\begin{equation}
    q(0) = Q_i, \p q(1) = Q_f, \p \frac{dq}{d\eps} = -\frac{\d\chi}{\d p}, \p \frac{dp}{d\eps} = \frac{\d\chi}{\d q}. \label{eq:eikpath}
\end{equation}
The path $q(\eps), p(\eps)$ is not the physical path $(Q(t), P(t))$ produced by the Hamiltonian $\h$. The two only generically coincide at the end-points. By \eqref{eq:hamexpdef}, the $(q(\eps), p(\eps))$ path is formally given by (that for any analytic $A(Q, P)$:
\begin{equation}
    A(q(\eps), p(\eps)) = e^{\eps\{\chi,\bt\}} A(Q_i, P_i). \label{eq:pathsolution}
\end{equation}

Continuing with the view that $-\chi$ is like an effective Hamiltonian, the on-shell action corresponding to $\chi$ is:
\begin{equation}
    \fc{I}(t_f, Q_f, t_i, Q_i) = \int_0^1\left(p(\eps)\frac{dq(\eps)}{d\eps} + \chi(t_f, t_i, q(\eps), p(\eps))\right)d\eps. \label{eq:eikfunc}
\end{equation}
We write that $\fc{I}$ depends on $t_f, t_i$ because $\chi$ does, but it is very important to understand that this is not transparently the same time dependence that an on-shell action normally has. The ``time'' corresponding to $\chi$ is $\eps$, and formally $\fc{I}$ could be written as a function of the initial and final values of $\eps$, but we always keep those equal to $0,1$, and so those hypothetical dependencies are not written. If we did carry them through, the initial and final $\eps$ derivatives of $\fc{I}$ would give the initial and final $\chi$ as usual by the Hamilton-Jacobi conditions. 

Let us now understand how $\fc{I}$ actually depends on its parameters. Under full variation of all of its arguments, we have:
\begin{align}
    \delta \fc{I} =&\ \left. p\delta q\right|^1_0 + \delta t_f\int_0^1\frac{\d \chi(t_f, t_i, q, p)}{\d t_f} d\eps + \delta t_i \int_0^1\frac{\d\chi(t_f, t_i, q, p)}{\d t_i} d\eps \nonumber \\
    &\ + \int_0^1\left(\delta p\left(\frac{dq}{d\eps} + \frac{\d\chi}{\d p}\right) - \delta q\left(\frac{dp}{d\eps} - \frac{\d\chi}{\d q}\right)\right)d\eps.
\end{align}
Applying that we use the Magnusian path \eqref{eq:eikpath}, this becomes:
\begin{equation}
    \delta \fc{I} = P_f\delta Q_f - P_i\delta Q_i + \delta t_f\int_0^1\frac{\d \chi}{\d t_f} d\eps + \delta t_i\int_0^1 \frac{\d\chi}{\d t_i} d\eps. 
\end{equation}
We can compute the remaining integrals by using the Magnus formulas \eqref{eq:magnusformula} and \eqref{eq:magnusinitial}. First, note that by \eqref{eq:pathsolution}:
\begin{align}
    \frac{\d\chi(t_f, t_i, Q_f, P_f)}{\d t_f} &= \frac{\d\chi(t_f, t_i, q(1), p(1))}{\d t_f} = e^{(1-\eps)\{\chi,\bt\}} \frac{\d\chi(t_f, t_i, q(\eps), p(\eps))}{\d t_f} \\
    \frac{\d\chi(t_f, t_i, Q_i, P_i)}{\d t_i} &= \frac{\d\chi(t_f, t_i, q(0), p(0))}{\d t_i} = e^{-\eps\{\chi,\bt\}} \frac{\d\chi(t_f, t_i, q(\eps), p(\eps))}{\d t_i}. 
\end{align}
The first of these, fed into \eqref{eq:magnusformula}, results in:
\begin{align}
    \int_0^1\frac{\d\chi(t_f, t_i, q(\eps), p(\eps))}{\d t_f} d\eps =&\ \int_0^1 e^{(1-\eps)\{\bt,\chi\}} d\eps \frac{\d\chi(t_f, t_i, Q_f, P_f)}{\d t_f} \nonumber \\
    =&\ \frac{e^{\{\bt,\chi\}}-1}{\{\bt,\chi\}} \frac{\{\bt,\chi\}}{e^{\{\bt,\chi\}}-1}(-\h(t_f, Q_f, P_f)) = -\h_f
\end{align}
while the second, following the same steps but with \eqref{eq:magnusinitial}, gives:
\begin{equation}
    \int_0^1\frac{\d\chi(t_f, t_i, q(\eps), p(\eps))}{\d t_i} d\eps = \h(t_i, Q_i, P_i) = \h_i.
\end{equation}
Therefore, the exact differential of $\fc{I}(t_f, Q_f, t_i, Q_i)$ is:
\begin{equation}
    d\fc{I} = P_f dQ_f - P_i dQ_i - \h_f dt_f + \h_i dt_i.
\end{equation}
$\fc{I}$ precisely satisfies all of the Hamilton-Jacobi conditions \eqref{eq:hamjac} and owing to \eqref{eq:chiinit} vanishes when $t_f = t_i$. Thus, due to the uniqueness of solutions to Hamilton's equations of motion, $\fc{I}$ must \emph{exactly} be the on-shell action corresponding to the true physical path $(Q(t), P(t))$ evolved with the true Hamiltonian! That is:
\begin{equation}
    \s(t_f, Q_f, t_i, Q_i) = \int_{t_i}^{t_f}\left(P \frac{dQ}{dt} - \h(t, Q, P)\right) dt = \int_0^1\left(p \frac{d q}{d\varepsilon} + \chi(t_f, t_i, q, p)\right) d\varepsilon \label{eq:eik2osa}
\end{equation}
where $(Q(t), P(t))$ is the physical path through phase space in time satisfying Hamilton's equations \eqref{eq:hamilton} and $(q(\eps), p(\eps))$ is the path satisfying \eqref{eq:eikpath} with the same boundary conditions as the physical path. We have therefore derived equation \eqref{eq:eik2osa} as an exact, fully general relationship between the Magnusian and the on-shell action.\footnote{Shortly after the submission of this paper, Ref.~\cite{Kim:2025sey} appeared where \eqref{eq:eik2osa} was stated with a description of an outline for its derivation.} 

We can complete the $\eps$ integration in \eqref{eq:eik2osa} to arrive at a formal relation between the Magnusian and the OSA which explicitly only depends on the boundary conditions. By \eqref{eq:pathsolution}, the integrand of \eqref{eq:eikfunc} can be written as:
\begin{equation}
    p(\eps)\frac{dq}{d\eps}(\eps) + \chi(t_f, t_i, q(\eps),p(\eps)) = e^{\eps\{\chi,\bt\}}(P_{i}\{\chi,Q_i\}) + \chi(t_f, t_i, Q_i, P_i). 
\end{equation}
The $\eps$ integral in \eqref{eq:eik2osa} can then be computed to give the novel half of \eqref{eq:main}:
\begin{equation}
    \s(t_f, Q_f, t_i, Q_i) = \frac{e^{\{\chi,\bt\}}-1}{\{\chi,\bt\}}(P_{i}\{\chi,Q_i\}) + \chi(t_f, t_i, Q_i, P_i). \label{eq:eik2osaformal}
\end{equation}
Equation \eqref{eq:eik2osaformal} is formally exact, but difficult to apply in a meaningful sense except perturbatively. If $\chi$ is known to some order in perturbation theory, then everything on the right hand side can be computed to that same order, so that $\s$ is then found to the appropriate order. To express $\s$ properly, it must be written as a function of $Q_f$ which can be found to the relevant order from $\chi$ using \eqref{eq:eikexpdef} in terms of $(Q_i, P_i)$ and then inverted to that order so that the $P_i$ appearances in $\s$ can be eliminated. 

By applying the inverse operator to the differential operator acting on $P_{i}\{\chi, Q_i\}$ in \eqref{eq:eik2osaformal}, we have:
\begin{equation}
    \chi + P_{i}\{\chi, Q_i\} = \frac{\{\chi,\bt\}}{e^{\{\chi,\bt\}}-1} \s
\end{equation}
which now is directly in the form appropriate for the Magnus series. This expression can be used to perturbatively compute $\chi$. To make sense of a Poisson bracket acting on the on-shell action, use that $Q_f$ is an implicit function of $Q_i$ and $P_i$ which can be computed (perturbatively) using the Hamilton-Jacobi conditions.

In the case of an integrable system, it is possible to go to action-angle variables with angle coordinates $\theta$ and adiabatically invariant momenta $J$. In such a case, the Hamiltonian is only a function of the $J$, and consequently the Magnusian is only a function of them as well. Immediately then \eqref{eq:eikexpdef} trivializes so that:
\begin{equation}
    \theta_f = \theta_i - \frac{\d \chi}{\d J_i}, \p J_f = J_i.
\end{equation}
This further causes \eqref{eq:eik2osaformal} to simplify significantly so that:
\begin{equation}
    \s(t_f, \theta_f, t_i, \theta_i) = J_i(\theta_f - \theta_i) + \chi(t_f, t_i, J_i). \label{eq:integrablesimple}
\end{equation}
So, in the case of an integrable system expressed in action-angle variables, the on-shell action and the Magnusian are related by a simple Legendre transformation exchanging the action variable $J$ for the coordinate impulse $\theta_f - \theta_i$.

\subsection{Quantumly} \label{sec:quantum}

We will now prove \eqref{eq:eik2osa} from quantum mechanics, which allows closer contact to traditional amplitudes techniques expressed in terms of the quantum S-matrix. In quantum mechanics, the Magnusian (for finite time evolution) is very simply defined in terms of the time evolution operator $\hat U$ (which becomes the S-matrix in the interaction picture and in the long time evolution limit). As well, the matrix elements of the time evolution operator can be related, in the classical limit and by using the path integral, very directly to the classical on-shell action. So, it is the path integral which will provide us the direct link between the action and the Magnusian. To appreciate this, we first need to carefully review the interplay between the classical limit and the path integral. 

Consider our Hilbert space $\mathbb{H}$ to be the space of complex square-integrable functions of $d$ real variables $(Q^1, ..., Q^d)$. Let $\hat Q^a$ be the usual coordinate operators on this space and $|Q\rangle$ be the corresponding ``coordinate-eigenstates''. Let $\hat P_a$ be the usual momentum operators conjugate to $\hat Q^a$ and let $|P\rangle$ be the corresponding ``momentum-eigenstates''. Assume we have chosen our coordinates so that $\<\hat Q^a\>$ and $\<\hat P_a\>$ have finite classical limits $Q^a_\cl$ and $P^\cl_a$ (in the sense of the limit defined by KMOC~\cite{Kosower:2018adc}). This classical-well-behaved-ness is automatic if our coordinates and momenta are the usual worldline coordinates and linear momenta of a finite collection of particles (massless or massive) or if they are the field operators of a massless field (which becomes a classical field in the classical limit of quantum field theory).\footnote{In the case of a massless quantum field theory, the coordinates can be the values of the fields at lattice points in a finite lattice with size regulated by $d$. The $d\to\8$ limit treated appropriately as usual will give the field theory limit.} As explained by KMOC, the classical limit (properly taken) of a massive quantum field is not a classical field theory but instead a collection of theories of different numbers of massive point-particles. Massive fields can be treated just fine by our description because within any given sector of particle numbers, the particles of that sector can be assigned classically finite worldline coordinates and linear momenta as usual, and in the classical limit the different sectors fully decouple.

Let $\hat G$ be an operator on our Hilbert space. From that operator, define the function $G(Q, P)$, called the inverse Weyl transform~\cite{LEE1995147, Moyal_1949} of $\hat G$, by:
	\begin{equation}
		G(Q, P) = \fc{W}^{-1}(\hat G) = \int e^{\frac{i}{\hbar}q^a P_a} \left\langle Q - \frac{q}{2}\right|\hat G\left|Q+\frac{q}{2}\right\rangle  d^d q.		\label{eq:wigner}
	\end{equation}
	Taking the conjugate of both sides shows that $G$ is a real-valued function if and only if $\hat G$ is Hermitian. We call $G$ the Wigner function associated to $\hat G$. Fourier transforming $G$ with respect to $P$ followed by a slight change of variables and use of the Baker-Campbell-Hausdorff formula allows one to invert the above and find that $\hat G$ is determined by $G$ as:
	\begin{equation}
		\hat G = \int e^{ik_a(\hat Q^a - Q^a) + i\ell^a(\hat P_a - P_a)} G(Q, P) \frac{d^d k d^d\ell d^d Q d^d P}{(2\pi)^{2d}}.	\label{eq:weyl}
	\end{equation}
	Very carefully note that this is not a quantization scheme. The definition of $G$ and its relationship to $\hat G$ make no use of or reference to any notion of classical limit and we have made no identification of $G$ with being the classical function ``corresponding'' to $\hat G$. In general such an identification is incorrect.\footnote{Even if you make this identification in one set of coordinates, under a change of coordinates the correspondence will be broken.} If $\hat G$ is such that $\<\hat G\>$ is a classically finite function $G_\cl(Q_\cl, P_\cl)$, then the Wigner function $G(Q, P)$ will have some $\hbar$ expansion:
    \begin{equation}
        G(Q, P) = \sum_{n=0}^\8 \hbar^n G_n(Q, P). \label{eq:wighbar}
    \end{equation}
    The classical limit will take $\hbar\to 0^+$ and then squeeze the wavefunction to localize around $(Q_\cl, P_\cl)$. Consequently:
    \begin{equation}
        G_0(Q, P) = G_\cl(Q, P). \label{eq:classicallimit}
    \end{equation}
    The details deriving this are given in Appendix \ref{sec:wigner}. Only the $\hbar$-independent piece of $G$ is fixed by the classical limit. Equation \eqref{eq:classicallimit} is the proper statement of how classical functions on phase space are contained within quantum operators. 

    The Wigner transform of a product is not the product of Wigner transforms. Instead, the induced product on Wigner functions is the star product $\star$, which becomes the ordinary product at leading order in $\hbar$:
    \begin{equation}
        \fc{W}^{-1}(\hat{\fc{W}}(A) \hat{\fc{W}}(B)) = A\star B = A e^{\frac{i\hbar}{2}\{\leftarrow,\rightarrow\}}B = A_0B_0 + \OO(\hbar) \label{eq:pbhbar}
    \end{equation}
    where $A_0$ and $B_0$ are the $\hbar\to 0^+$ limits of the Wigner functions $A$ and $B$. The leading order behavior ensures that in the classical limit the expectation of a product becomes the product of expectations. The notation $\{\leftarrow,\rightarrow\}$ means the bi-differential operator that is the classical Poisson bracket with the left-arrow/right-arrow indicating differentiation in that position of only functions to its left/right. Using the Weyl transform, the induced bracket on Wigner functions is the Moyal bracket $\mlb \bt, \bt \mrb$, and it immediately follows that:
    \begin{equation}
        \fc{W}^{-1}\left(\frac{1}{i\hbar}[\hat A, \hat B]\right) = \mlb A, B \mrb = A(Q,P) \frac{2}{\hbar}\sin\left(\frac{\hbar}{2}\{\leftarrow,\rightarrow\}\right)B(Q, P) = \{A_0, B_0\} + \OO(\hbar)
    \end{equation}
    which is the proper statement relating commutators and classical Poisson brackets. For more thorough coverage of Wigner functions, the Weyl transform, and the Moyal bracket, see Refs.~\cite{LEE1995147, Moyal_1949}.

    The path integral is most naturally expressed in terms of Wigner functions (and is likewise not a quantization scheme). Let $\hat U(t, t_i)$ be the usual time evolution operator and $U(t_f, Q_f, t_i, Q_i)$ be the usual propagation amplitude satisfying:
	\begin{equation}
		\hat U(t, t) = \mathbbm{1}, \p i\hbar\frac{\partial\hat U}{\partial t}(t, t_i) = \hat H(t)\hat U(t, t_i), \p U(t_f, Q_f, t_i, Q_i) = \langle Q_f|\hat U(t_f, t_i)|\hat Q_i\rangle \label{eq:Udef}
	\end{equation}
    with $\hat H(t)$ the Hamiltonian operator of the system. Define the functional $\s_H[Q,P]$ of the path $(Q(t'), P(t'))$ by:
	\begin{equation}
		\s_H[Q, P] = \int_{t_i}^{t}\Bigl(P_a(t') dQ^a(t') - H(t', Q(t'), P(t'))dt'\Bigr)
	\end{equation}
    where $H$ is the Wigner function associated to the Hamiltonian operator $\hat H$. The path integral measure $\fc{D}_\hbar Q \fc{D}_\hbar P$ is formally defined by:
    \begin{equation}
        \fc{D}_\hbar Q \fc{D}_\hbar P = \lim_{N\to\8} \frac{d^d P_0}{(2\pi\hbar)^d} \frac{d^d Q_1 d^d P_1}{(2\pi\hbar)^d} ... \frac{d^d Q_{N-1} d^d P_{N-1}}{(2\pi\hbar)^d}. \label{eq:pathintegralmeasure}
    \end{equation}
    The path integral is:
	\begin{align}
		U(t_f, Q_f, t_i, Q_i) &= \int_{Q(t_i) = Q_i}^{Q(t_f) = Q_f} e^{\frac{i}{\hbar}\s_H[Q,P]} \mathcal{D}_\hbar Q\mathcal{D}_\hbar P\label{eq:pathintegral}
	\end{align}
    where the path $Q(t), P(t)$ is broken up into pieces $Q(t_n) = Q_n$, $P(t_n) = P_n$ with $Q_0 = Q_i$, $Q_N = Q_f$, $t_n = t_i + \frac{n}{N}(t_f-t_i)$ and the Wigner action functional $\s_H$ is understood through a Riemann sum. The path integral in this form is straightforwardly derived from the short-time evolution behavior of $\hat U$, combined with its composition rule $\hat U(t_f, t_i) = \hat U(t_f, t')\hat U(t',t_i)$, and further combined with the expression \eqref{eq:weyl} for the function $H$. $H$ cannot be generally replaced by $\mathcal{H}_\text{cl}$. Instead, by \eqref{eq:classicallimit}, $\mathcal{H}_\text{cl} = \lim_{\hbar\to 0^+} H$ and a generic quantum theory will have $\mathcal{O}(\hbar)$ corrections between the two. Such corrections may be strictly necessary in order to have the correct symmetries as a quantum theory. 

    The path integral provides a direct way to compute the leading classical behavior of the matrix elements of the time evolution operator. In the $\hbar\to 0^+$ limit, it becomes a stationary phase integral with the relevant contribution coming precisely from the classical trajectory. Thus, we expect that $U(t_f, Q_f, t_i, Q_i)$ should contain a prefactor $e^{\frac{i}{\hbar}\s(t_f, Q_f, t_i, Q_i)}$ with $\s(t_f, Q_f, t_i, Q_i)$ the classical on-shell action. As well, the integral measure of the path integral contains many powers of $\hbar$ (formally infinite), and it is not clear a priori how many of these factors will survive in the classical limit. Direct calculation shows that almost all of those powers of $\hbar$ are canceled in the stationary phase integration, so that the leading order classical behavior of the path integral is:
    \begin{equation}
        \lim_{\hbar\to 0^+} \hbar^\frac{d}{2} e^{-\frac{i}{\hbar} \s(t_f, Q_f, t_i, Q_i)} U(t_f, Q_f, t_i, Q_i) = \left(\frac{-1}{2\pi i}\right)^{\frac{d}{2}}\sqrt{\text{det} \frac{\partial^2 \s}{\partial Q_f \partial Q_i}} e^{-i\int_{t_i}^{t_f} H_1 dt}. \label{eq:leadingclassical}
    \end{equation}
    The details deriving this are given in Appendix \ref{sec:quant}. 

    From \eqref{eq:leadingclassical}, it immediately follows that the classical on-shell action can be extracted from the logarithm of the matrix elements of the time evolution oeprator as:
    \begin{equation}
        i\s(t_f, Q_f, t_i, Q_i) = \lim_{\hbar\to 0^+} \hbar\ln U(t_f, Q_f, t_i, Q_i).
    \end{equation}
    In the case of scattering, the interaction picture is used, the initial and final times are taken to $\pm\8$, and the time evolution operator becomes the S-matrix. The resulting interaction picture on-shell action we write as $\s(Q_f, Q_i)$ (in this case the coordinates must be selected to have well-defined values in the $t_{f/i} \to \pm\8$ limits). Comparing to \eqref{eq:deltadef}, we find the left half of \eqref{eq:main}:
    \begin{equation}
        \delta = \s(Q_f, Q_i)
    \end{equation}
    which confirms the well-known result that the classical behavior of the optical eikonal phase is just the on-shell action. 

The time-evolution operator, by virtue of being unitary, can be written in the form:
\begin{equation}
    \hat U(t_f, t_i) = e^{\frac{i}{\hbar}\hat\chi(t_f,t_i)} \label{eq:expophase}
\end{equation}
where $\hat\chi$ is the Hermitian eikonal operator. Let $\chi_{qm}$ be its inverse Weyl transform and let $\chi$, the classical eikonal generator (Magnusian) be: 
\begin{equation}
    \chi(t_f, t_i, Q, P) = \lim_{\hbar\to 0^+}\chi_\text{qm}(t_f, t_i, Q, P).
\end{equation}
Consider a quantum system with initial wavefunction $|\psi_i\rangle$ to undergo time evolution to some final state $|\psi_f\rangle$. Let $\hat A$ be a Hermitian operator so that $\langle\hat A\rangle$ is classically finite and let $A(Q, P)$ be the classical limit of its Wigner function. Then, the final value of $\hat A$ is given by:
\begin{equation}
    \langle\hat A\rangle_f = \langle\psi_f|\hat A|\psi_f\rangle = \langle\psi_i|\hat U^\dagger \hat A \hat U|\psi_i\rangle = \langle\psi_i| e^{-\frac{i}{\hbar}\hat\chi}\hat A e^{\frac{i}{\hbar}\hat\chi}|\psi_i\rangle = \langle\psi_i|\left(e^{\frac{1}{i\hbar}[\hat\chi,\bt]} \hat A\right)|\psi_i\rangle. \label{eq:qmexp}
\end{equation}
Using \eqref{eq:pbhbar}, we recover \eqref{eq:eikexpdef} in the classical limit. There is no subtlety of $\hbar$ over $\hbar$ cancellations in connecting \eqref{eq:qmexp} and \eqref{eq:eikexpdef} because the commutator $[\bt, \bt]$ of any operators with classically finite expectation values is of $\OO(\hbar)$, so $\frac{1}{\hbar}[\bt, \bt]$ contains no inverse powers of $\hbar$.

From \eqref{eq:expophase} and \eqref{eq:leadingclassical} we can explicitly relate the Magnusian and the on-shell action. Let $\hat V(\varepsilon)$ be the  1-parameter family of unitary operators defined by $\hat V(\eps) = e^{\frac{i}{\hbar}\varepsilon \hat \chi}$. This $\hat V$ immediately satisfies \eqref{eq:Udef} with $t$ replaced by $\varepsilon$, $t_i$ replaced by $0$, $\hat H(t)$ replaced by $-\hat\chi$, and $\hat U$ replaced by $\hat V$. As well, $\hat V(1) = \hat U(t_f, t_i)$. Thus, from the path integral it is immediately true that exactly:
\begin{align}
		U(t_f, Q_f, t_i, Q_i) &= \int_{Q(0) = Q_i}^{Q(1) = Q_f} e^{\frac{i}{\hbar}\int_0^1\left(P_a\frac{dQ^a}{d\varepsilon} +\chi_\text{qm}(t_f, t_i, Q, P)\right) d\varepsilon} \mathcal{D}_\hbar Q\mathcal{D}_\hbar P. 
\end{align}

We can now study the leading classical behavior of the above expression using the same techniques as previously. In particular, let $(q(\varepsilon), p(\varepsilon))$ again be the path satisfying \eqref{eq:eikpath} and let $\fc{I}$ again be given by \eqref{eq:eikfunc}. Then, in the classical limit, the same calculation that led to \eqref{eq:leadingclassical} gives:
\begin{equation}
    \lim_{\hbar\to 0^+} \hbar^\frac{d}{2} e^{-\frac{i}{\hbar}\fc{I}} U(t_f, Q_f, t_i, Q_i) = \left(\frac{-1}{2\pi i}\right)^\frac{d}{2} \sqrt{\text{det}\frac{\partial^2 \fc{I}}{\partial Q_f\partial Q_i}} e^{i\int_0^1 \chi_1 d\varepsilon}.
\end{equation}
In order for this and \eqref{eq:leadingclassical} to be true, because both $\s$ and $\chi$ are $\hbar$ independent, the $\frac{1}{\hbar}$-singular exponentials $e^{\frac{i}{\hbar}\s}$ and $e^{\frac{i}{\hbar}\fc{I}}$ must match, giving:
\begin{equation}
    \s(t_f, Q_f, t_i, Q_i) = \int_0^1\left(p_a \frac{d q^a}{d\varepsilon} + \chi(t_f, t_i, q, p)\right) d\varepsilon
\end{equation}
exactly reproducing \eqref{eq:eik2osa}. 

One perspective on the origin of the complicated relationship between $\delta$ ($=\s$) as defined in \eqref{eq:deltadef} and $\chi$ as defined in \eqref{eq:chidef} is via $\hbar$ over $\hbar$ cancellations. The issue is ultimately that the exponential of a Weyl transform is not the Weyl transform of the exponential:
\begin{equation}
    \hat{\text{S}} = e^{\frac{i}{\hbar}\hat{\fc{W}}(\chi_\text{qm})} \neq \hat{\fc{W}}(e^{\frac{i}{\hbar}\chi_\text{qm}}).
\end{equation}
In particular, the operator exponential when projected into phase space is naturally expressed in terms of the star product:
\begin{equation}
    \fc{W}^{-1}(e^{\frac{i}{\hbar}\hat{\fc{W}}(\chi_\text{qm})}) = 1 + \frac{i}{\hbar}\chi_\qm - \frac{1}{2\hbar^2}\chi_\qm\star\chi_\qm - \frac{i}{6\hbar^3}\chi_\qm\star\chi_\qm\star\chi_\qm + \cdots
\end{equation}
Each star product, via \eqref{eq:pbhbar}, can be expanded as a series in $\hbar$, and plainly the term containing $n$ powers of $\chi_\qm$ must be computed up to order $\hbar^{n-1}$ due to the presence of a $\frac{1}{\hbar^n}$ denominator in order to account for all terms that will get resummed into the form $e^{\frac{i}{\hbar}\delta}$. We have explicitly checked through the fourth order in powers of $\chi$ that doing so gives precisely the expression necessary to resum to \eqref{eq:eik2osaformal}. 

\section{Properties of the Magnusian}\label{sec:properties}

\subsection{Interpretation}

Although the Magnusian has been of increasingly popular interest in amplitudes literature in recent years, we can see from the derivation of \eqref{eq:eik2osaformal} that in fact its existence has nothing to do with the context of scattering or the consideration of any particular theory. The Magnusian is a generic quantity of interest in classical mechanics that captures, through \eqref{eq:eikexpdef}, the entire time evolution of the system. Equation \eqref{eq:eik2osa} gives us perspective on what the Magnusian is actually doing. Given a generically time dependent true Hamiltonian (as is the standard case in the interaction picture), the resultant physical path through phase space tends to be very complicated. The Magnusian acts as a sort of resummed/time-averaged Hamiltonian so that the path it gives through phase space (when treated as an effective Hamiltonian with the same boundary conditions as the true path) is a simple exponential \eqref{eq:pathsolution}. To make this precise, notice that by comparing \eqref{eq:eikexpdef} to \eqref{eq:hamexpdef}, which is valid for a time independent Hamiltonian $\h$ and interval of time evolution $[\eps_i, \eps_f]$, we can define the time-averaged Hamiltonian:
\begin{equation}
    \bar{\h}_T(t, Q, P) = -\frac{1}{T}\chi(t+T, t, Q, P)
\end{equation}
and if the system is evolved from initial conditions $(Q_i, P_i)$ using this Hamiltonian $\bar{\h_T}(t, Q, P)$ (rather than the physical Hamiltonian $\h(t, Q, P)$) from an initial time $t$, then at final time $t+T$ it will have the physically correct final conditions $(Q_f, P_f)$. By virtue of the time integrals over the $[t, t+T]$ interval which must be completed to compute $\chi$ from the physical Hamiltonian $\h$ (by following the Magnus formula \eqref{eq:magnusformula}), the high-frequency behavior (compared to $\frac{1}{T}$) of the Hamiltonian is effectively integrated out. Using the Magnus series to express $\bar{\h}_T$ in terms of $\h$ gives that the Magnusian path evolved using $\bar{\h}_T$ is a ``near-identity transformation'' of the true physical path evolved using $\h$~\cite{nit}.

The fact that the time-evolution generated by $\bar{\h}_T$ is a simple exponential in the space of canonical transformations becomes geometrically meaningful when we think about that space as a Lie group. A Lie group, viewed as a manifold and given a connection made from only the structure constants of the algebra, has geodesics (parameterized by affine parameter $\eps$) which are precisely exponentials $e^{\eps X}$ with $X$ an element of the Lie algebra. Thus, the Magnusian essentially computes the geodesic through the group of canonical transformations which connects the identity to the canonical transformation provided by physical time evolution. That geodesic path as a group element is $e^{\eps\{\chi,\bt\}}$. 

\begin{figure}[t!]
\centering
        \includegraphics[totalheight=8cm]{
        %example_affinegroup.png
        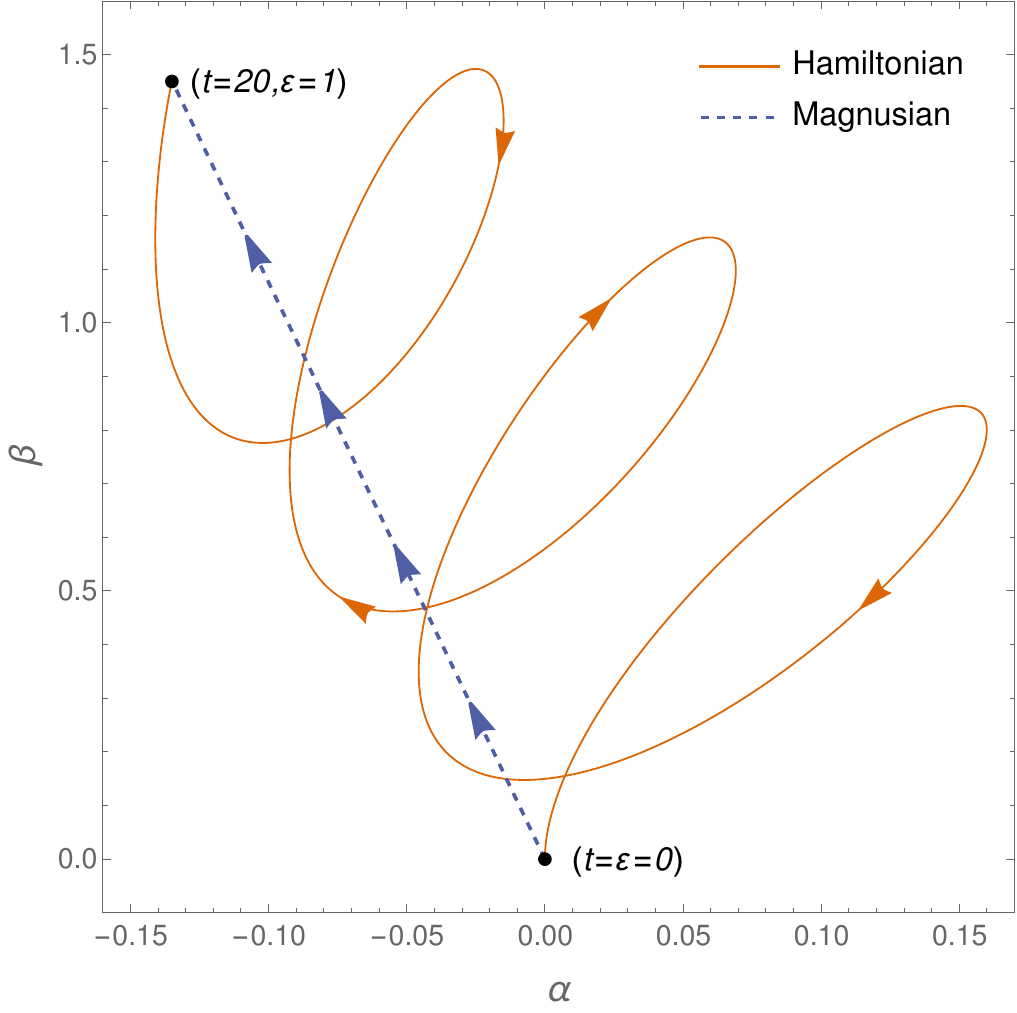
        }
    \caption{Example of group trajectories for affine transformations due to evolution using the physical Hamiltonian (shown in solid orange) and Magnusian (shown in dashed blue). This is computed for the case of $u(t)=0.1 \sin(t)$ and $v(t)=0.05+0.3 \cos(t)+0.3 \sin(t)$. 
    }
    \label{fig:affinegroup}
\end{figure}

We can make the sense in which that path is properly geodesic more concrete by looking at a low dimensional example. On a generic phase space, even of only one coordinate and one momentum, the group of canonical transformations is infinite dimensional. But, if we consider only transformations preserving some natural structure, we can produce simple finite dimensional cases. The smallest nontrivial possible example is the affine group over the real numbers. That is, the group of transformations of the form $x' = e^\beta x + \alpha$ with parameters $\alpha,\beta$ giving the amounts of translation and dilation respectively. We write $G(\alpha,\beta)$ for a general such element of the group. We can identify a matrix representation for the group by defining the generator matrices $\hat \tau_1$ and $\hat \tau_2$:
\begin{equation}
    \hat \tau_1 = \begin{pmatrix} 0 & -1 \\ 0 & 0 \end{pmatrix}, \ps \hat \tau_2 = \begin{pmatrix} 1 & 0 \\ 0 & 0 \end{pmatrix}\ \ \implies \ \
    [\hat \tau_1, \hat \tau_2] = -\hat \tau_1, \ps e^{\alpha\hat \tau_1}e^{\beta\hat \tau_2}\begin{pmatrix} x \\ -1\end{pmatrix} = \begin{pmatrix} e^\beta x + \alpha\\ -1 \end{pmatrix}. 
\end{equation}
From a matrix representation of a Lie group it is always possible to construct a classical mechanical system that represents that group through functions on phase space that are bilinear in $(Q, P)$. In particular, consider a 4 dimensional phase space $(Q_1, Q_2, P_1, P_2)$ and define:
\begin{equation}
    \tau_1 = P_a(\hat \tau_1)^a{}_b Q^b = -P_1 Q_2, \ps \tau_2 = P_a(\hat \tau_2)^a{}_b Q^b = P_1 Q_1 \ \ \implies \ \ \{\tau_1, \tau_2\} = -\tau_1. 
\end{equation}
We can give arbitrary dynamics on the group by considering a Hamiltonian that is a generic element of the Lie algebra at each time:
\begin{equation}
    \h(t, Q, P) = u(t) \tau_1 + v(t) \tau_2
\end{equation}
for some functions $u(t)$ and $v(t)$. The equations of motion produced by $\h$ are easily exactly solved, and give that $(Q(t), P(t))$ is related to $(Q(0), P(0))$ by the action of the group element $G(\alpha(t), \beta(t))$ with the physical group trajectory given by:
\begin{equation}
    \alpha(t) = \frac{\beta(t)}{1-e^{-\beta(t)}} \int_0^t u(t') e^{-\beta(t')} dt', \p  \beta(t) = \int_0^t v(t') dt'.
\end{equation}
Solving the $\eps$-equations of motion produced by $\chi$ and requiring a match at $\eps = 1$ to $t = T$ gives that the Magnusian trajectory is:
\begin{equation}
    \alpha_\chi(\eps) = \eps\alpha(T), \p \beta_\chi(\eps) = \eps\beta(T). 
\end{equation}
Suppose that we have a system for which $u$ and $v$ are naturally oscillatory, which could arise (in the interaction picture) by coupling to some external field of a given frequency $\omega$. A visualization of the evolution of parameters $(\alpha,\beta)$ in such a case is shown in figure \ref{fig:affinegroup}.

The next simplest example after the affine group is motion on $SU(2)$, a similar analysis and visualization of which are given in appendix~\ref{sec:su2}. 

\subsection{Symmetries}

Let us understand the symmetries of the Magnusian. Consider a generic continuous canonical transformation. That is, some map from old $t, Q, P, \h$ into new $t', Q', P', \h'$ so that Hamilton's equations are satisfied in the primed quantities if and only if they are satisfied in the unprimed quantities, and so that this map is smooth with respect to a parameter $\alpha$ with $\alpha = 0$ giving the identity transformation. Define the functions $\delta t(t, Q, P)$ and $\delta Q(t, Q, P)$ by:
\begin{equation}
    \delta t(t, Q, P) = \left.\frac{\d t'(\alpha, t, Q, P)}{\d\alpha}\right|_{\alpha = 0}, \p \delta Q(t, Q, P) = \left.\frac{\d Q'(\alpha, t, Q, P)}{\d\alpha}\right|_{\alpha = 0}.
\end{equation}
In order for such a continuous transformation to be canonical (and so preserve the Poisson bracket), there must exist a phase space function $G(t, Q, P)$ called the generator of the transformation so that for any fixed function $A(Q, P)$ on phase space:
\begin{equation}
    \left.\frac{d A(Q'(t), P'(t))}{d\alpha}\right|_{\alpha = 0} = \{A, G\}.
\end{equation}
Under such a canonical transformation, the new Hamiltonian $\h'(\alpha, t', Q', P')$ will generally have a different functional form than the untransformed Hamiltonian. We write $\Delta\h$ to represent the variation in the functional form of the Hamiltonian at the identity. By the chain rule, this is related to the total $\alpha$ derivative (in which its arguments $Q', P'$ are also varied) by:
\begin{equation}
    \left.\frac{d\h'(\alpha, t', Q'(t'), P'(t'))}{d\alpha}\right|_{\alpha = 0} = \Delta\h + \frac{\d\h}{\d t}\delta t + \{\h, G\}. 
\end{equation}
By Noether's theorem, the change in functional form of the Hamiltonian exactly gives the rate of nonconservation of the generator. That is:
\begin{equation}
    \frac{dG}{dt} = \Delta \h
\end{equation}
so that the transformation is a symmetry if and only if its generator is conserved. 

Similarly to the Hamiltonian, the transformed on-shell action $\s'(\alpha, t_f', Q'_f, t_i', Q'_i)$ will generally have a different functional form than its untransformed version. We again write $\Delta \s$ to represent the variation in that functional form at the identity, and by the chain rule and Hamilton-Jacobi conditions \eqref{eq:hamjac}:
\begin{equation}
    \left.\frac{d\s'}{d\alpha}\right|_{\alpha = 0} = \Delta \s + P_f\delta Q_f - P_i \delta Q_i - \h_f \delta t_f + \h_i \delta t_i. 
\end{equation}
By comparing this to \eqref{eq:actvariation}, we immediately have that the changes in functional form of the action and the Hamiltonian are related by:
\begin{equation}
    \Delta \s = -\int_{t_i}^{t_f} \Delta \h dt.
\end{equation}
Due to Noether's theorem, this ultimately simplifies to:
\begin{equation}
    \Delta \s = G(t_i, Q_i, P_i) - G(t_f, Q_f, P_f). \label{eq:actfunctionalchange}
\end{equation}
The functional form of the action varies precisely by (negative of) the total nonconservation of the transformation generator. 

Following identical steps but instead beginning from the right hand side of \eqref{eq:eik2osa}, we find that the change in functional form of the Magnusian is related to that of the OSA by:
\begin{equation}
    \Delta \s = \int_0^1 \Delta\chi(t_f, t_i, q(\eps), p(\eps))d\eps.
\end{equation}
Using \eqref{eq:pathsolution}, the integration can be completed to give:
\begin{equation}
    \Delta \s = \frac{e^{\{\chi,\bt\}}-1}{\{\chi,\bt\}} \Delta\chi(t_f, t_i, Q_i, P_i). \label{eq:eikfunctionalchange}
\end{equation}
Manifestly, a symmetry of the action (a transformation for which the left side is 0) is automatically a symmetry of the Magnusian (for which the right is zero), and vice versa. That is, the Magnusian and the action have precisely the same symmetries.

By solving for the functional change of the Magnusian and using \eqref{eq:actfunctionalchange}, we can rewrite \eqref{eq:eikfunctionalchange} as:
\begin{equation}
    \Delta\chi(t_f, t_i, Q, P) = \{G(t_f, Q, P), \chi\} - \frac{\{\chi,\bt\}}{e^{\{\chi,\bt\}} -1}(G(t_f, Q, P) - G(t_i, Q, P)). 
\end{equation}
For most transformations of physical interest the generator has no explicit time dependence, in which case only the first term on the right hand side is nonvanishing and so the change in the Magnusian under such a transformation is given simply in terms of its Poisson bracket with the generator of that transformation. 

For scattering applications, a particularly important class of transformations is the class of field redefinitions which become the identity at infinity. Suppose we perform such a field redefinition. Because it asymptotes to the identity transformation, the generator $G$ of the transformation asymptotes to 0. Then, by \eqref{eq:actfunctionalchange}, the on-shell action is unsurprisingly unchanged by the transformation. It immediately follows from \eqref{eq:eikfunctionalchange} that the Magnusian is also unchanged by the transformation. Therefore, in the scattering case, the Magnusian is field redefinition invariant.

\subsection{Perturbative Relations}

Let us now see how \eqref{eq:eik2osa} can be used to compute the Magnusian from the action or vice versa in perturbation theory. For the rest of this section, we assume that we have already gone to the interaction picture and we hold $t_f, t_i$ fixed (in the scattering context, they are taken to $\pm\8$). Because $t_f, t_i$ are held fixed, we will not write them explicitly as arguments of functions which in principle depend on them. Because the Hamiltonian is $\OO(\lambda)$, the velocity $\dot Q$ and consequently the Lagrangian and then the action are also of $\OO(\lambda)$. By the Hamilton-Jacobi conditions \eqref{eq:hamjac}, the OSA is a natural function of the boundary conditions $Q_f$, $Q_i$. To see how this manifests in perturbation theory, we give $Q_f$ and $P_f$ perturbative expansions as:
\begin{equation}
    Q_f = \sum_{n=0}^\8 \lambda^n Q_n, \p P_f = \sum_{n=0}^\8 \lambda^n P_n
\end{equation}
where $Q_0 = Q_i$ and $P_0 = P_i$. Expanding the Hamilton-Jacobi differential \eqref{eq:hamjacdiff} (with $dt_f = dt_i = 0$) in powers of $\lambda$ gives:
\begin{equation}
    d\s = \sum_{n=1}^\8 \lambda^n \sum_{k=0}^n P_{(n-k)} dQ_k.
\end{equation}
Consequently, $\s$ is naturally written as a series in the form:
\begin{equation}
    \s(Q_f, Q_i) = \sum_{n=1}^\8 \lambda^n \s_n(Q_n, Q_{n-1}, ..., Q_1, Q_0)
\end{equation}
where the perturbation $\s_n$ satisfy the perturbative Hamilton-Jacobi conditions:
\begin{equation}
    P_{(n-k)} = \frac{\d \s_n}{\d Q_k}. \label{eq:perthamjac}
\end{equation}
The $k=0$ condition provides new information (the value of the perturbation $P_{n}$), while the $k > 0$ conditions determine derivatives of $\s_n$ in terms of already known derivatives of $\s_{n-k}$. 

The Magnusian also is of $\OO(\lambda)$ and so we give it a perturbative expansion in $\lambda$ as:
\begin{equation}
    \chi(Q, P) = \sum_{n=1}^\8 \lambda^n \chi_n(Q, P). 
\end{equation}
We can now expand \eqref{eq:eik2osaformal} directly in powers of $\lambda$. At $\OO(\lambda)$, we find:
\begin{equation}
    \s_1(Q_1, Q_i) = P_{i} Q_1 + \chi_1(Q_i, P_i)
\end{equation}
and so $\s_1$ is indeed an ordinary Legendre transform of $\chi_1$, interchanging $P_i$ and $Q_1$. 

At $\OO(\lambda^2)$, we find:
\begin{equation}
    \s_2(Q_2, Q_1, Q_i) = \chi_2(Q_i, P_i) + P_i Q_2 + \frac{1}{2} P_1 Q_1.
\end{equation}
This is no longer a simple Legendre transformation. However, given $\s_1$ and $\s_2$, all of $Q_1, P_1, Q_2$ can written in terms of $Q_i, P_i$ using the Hamilton-Jacobi conditions \eqref{eq:perthamjac}, allowing the determination of $\chi_2$. Expanding \eqref{eq:eik2osaformal} to $\OO(\lambda^3)$ gives:
\begin{equation}
    \s_3 = \chi_3 + P_i Q_3 + \frac{1}{2}P_2 Q_1 + \frac{1}{2} P_1 Q_2 + \frac{1}{12}P_1\{\chi_1,Q_1\} - \frac{1}{12}Q_1\{\chi_1,P_1\}.
\end{equation}
Given $\s_1, \s_2, \s_3$, every piece in this equation aside from $\chi_3$ can be computed in terms of $Q_i, P_i$ using \eqref{eq:perthamjac}, allowing the determination of $\chi_3$. Plainly based on what terms are equated at each order in $\lambda$ in \eqref{eq:eik2osaformal}, this pattern continues at higher orders, so that the Magnusian at $\OO(\lambda^n)$ can be determined from the $\s_k$ for $k \le n$. The converse is also true. Given the Magnusian up through order $\OO(\lambda^n)$, the on-shell action up to that same order can be computed by matching at each order in \eqref{eq:eik2osaformal} and integrating the conditions \eqref{eq:perthamjac} to the correct variable dependencies.

\section{Examples and Applications}\label{sec:exandapp}

\subsection{Simple Harmonic Oscillator} \label{sec:SHOex}
The simplest example of an exactly solvable system is the simple harmonic oscillator. Since this is an integrable system, the relation between the Magnusian and the on-shell action becomes just a Legendre transformation when considered in action-angle variables. Instead in this section we will treat the potential term perturbatively and show that indeed in this case the Magnusian and the on-shell action are non-trivially different and that they are related exactly by \eqref{eq:eik2osa}. 

The Hamiltonian for the simple harmonic oscillator is given as,
\begin{equation}
    \h(t, q,p) = \frac{p^2}{2} + \lambda \frac{q^2}{2} = \h_0 + \lambda \h_{I} \,.
\end{equation}
For this the exact solutions are
\begin{equation}
    q(t) = Q_i \cos \left(\sqrt{\lambda} t\right) + \frac{P_i}{\sqrt{\lambda} } \sin \left(\sqrt{\lambda} t \right) \,, \p 
    p(t) = P_i \cos \left(\sqrt{\lambda} t\right) - \sqrt{\lambda} Q_i \sin \left(\sqrt{\lambda} t \right) \,,  \label{eq:freeos_solqp}
\end{equation}
where $Q_i$ and $P_i$ are initial conditions at $t = 0$. Now we consider the case where $\lambda$ is small so that we take the kinetic term as the free Hamiltonian $\h_0$ and the potential term as the interaction part $\h_{I}$. We can then go to the interaction picture by a change of variables $(q,p)\rightarrow (Q,P)$ using the relations,
\begin{align}\label{eq:freeos_QPqp}
    Q (t) = q(t) - p(t) t \,, \quad P (t) = p(t) \,,
\end{align}
which are achieved by the generating function $-\s_0(t, q, 0, Q) = -\frac{(q-Q)^2}{2t}$. This generating function takes us to the new Hamiltonian $\h'$ given by
\begin{align}
    \h' = \h + \frac{\partial \s_0}{\partial t} = \lambda \frac{(Q + Pt)^2}{2} \,.
\end{align}
We can then compute the on-shell action as a perturbative series in $\lambda$ from the exact solutions of $(Q,P)$ using eqs. \eqref{eq:freeos_solqp} and \eqref{eq:freeos_QPqp}. The on-shell action up to $\mathcal{O}(\lambda^4)$ is given by,
\begin{align}
\begin{aligned}
    \s (T,0,Q_i, P_i) &= \int_0^{T} \left[ P(t) \frac{dQ(t)}{dt} - \h' (t, Q(t),P(t)) \right] dt
   \\ &= \frac{1}{6} \lambda  \left(P_i^2 T^3-3 Q_i^2 T\right) -\frac{1}{30} \lambda ^2 T^3 \left(3 P_i^2 T^2+10 P_i Q_i T+5
   Q_i^2\right)
   \\ &\phantom{=asdfasdfasdf} +\frac{1}{630} \lambda ^3 T^5 \left(10
   P_i^2 T^2+56 P_i Q_i T+63 Q_i^2\right) + \mathcal{O} (\lambda^4) \,.
\end{aligned} \label{eq:SHO_OSA}
\end{align}
We can also compute the Magnusian by iteratively integrating the equation \eqref{eq:magnusseries}. The result for the Magnusian up to $\mathcal{O}(\lambda^4)$ is given by,\footnote{As a check, we also use this Magnusian to time evolve $Q_i$ and $P_i$ using equation \eqref{eq:eikexpdef} and then compare it to the exact solution using eq. \eqref{eq:freeos_solqp}.}
\begin{align}
\begin{aligned}
    \chi (T,0,Q_i,P_i) &= -\frac{1}{6} \lambda  T \left(P_i^2 T^2+3 P_i Q_i T+3 Q_i^2\right) +\frac{1}{60} \lambda ^2 T^3 \left(P_i^2 T^2+5 P_i Q_i T+5 Q_i^2\right)
   \\ &\phantom{=asdfasdfasdfasdf}-\frac{\lambda ^3 T^5 \left(11
   P_i^2 T^2+42 P_i Q_i T+42 Q_i^2\right)}{3780} + \mathcal{O} (\lambda^4) \,.
\end{aligned}
\end{align}
The above Magnusian clearly does not agree with the on-shell action given in equation \eqref{eq:SHO_OSA}. But, a direct check confirms that the two are related by \eqref{eq:eik2osa}. To verify this first we need the path $(q(\eps),p(\eps))$ as a function of the ``$\varepsilon$-time''. Since in the interaction picture the Magnusian starts at $\mathcal{O}(\lambda)$, we can integrate eq. \eqref{eq:qpfrommagnusian} order by order to obtain,
 \begin{align}
    q(\varepsilon) &= Q_i + \frac{T^2 (3 Q_i + 2 P_i T) \varepsilon}{6} \lambda - 
 \frac{ T^4 \varepsilon (4 P_i T + 5 Q_i (2 + \varepsilon))}{120} \lambda^2 \nonumber\\
  &\quad\quad\quad+ \frac{T^6 \varepsilon (2 P_i T (44 - 35 \varepsilon^2) -  21 Q_i (-8 - 2 \varepsilon + 5 \varepsilon^2)) }{15120} \lambda^3 + \mathcal{O} (\lambda^4)\\
    p(\varepsilon) &= P_i - \frac{T (2 Q_i + P_i T)\varepsilon}{2}   \lambda + \frac{T^3 (4 Q_i - P_i T (-2 + \varepsilon)) \varepsilon}{24}  \lambda^2 \nonumber\\
  &\quad\quad\quad+ \frac{T^5 \varepsilon (2 Q_i (-8 + 5 \varepsilon^2) + P_i T (-8 + 2\varepsilon + 5 \varepsilon^2)) }{720} \lambda^3 + \mathcal{O} (\lambda^4).
\end{align}
Using the above we can now compute the relation between the on-shell action and the Magnusian as,
\begin{align}
\begin{aligned}
    \int_0^1 p \frac{d q}{d\varepsilon} d\eps &= \frac{P_iT^2(3Q_i + 2P_iT)\lambda}{6} - \frac{T^3(15Q_i^2 + 25P_iQ_iT + 7P_i^2T^2)\lambda^2}{60} \\
    &\quad\quad\quad+ \frac{T^5(420Q_i^2 + 378P_iQ_iT + 71P_i^2T^2)\lambda^3}{3780} + \mathcal{O}(\lambda^4)
\end{aligned}
\end{align}
which is exactly the difference between the Magnusian and the on-shell action. We have verified the relation in eq. \eqref{eq:eik2osa} up to $\mathcal{O}(\lambda^{10})$. The results can be found in the attached Mathematica notebook with this article.

\subsection{Anharmonic Oscillator}
The next simplest example we consider is a quartic perturbation to the harmonic oscillator. The Hamiltonian for the anharmonic case is given as
\begin{align}
    \h(t, q,p) = \frac{p^2}{2} + \frac{q^2}{2} + \lambda  \frac{q^4}{4}= \h_0 + \lambda \h_{I} \,,
\end{align}
where the free part part of the system is the harmonic oscillator and the quartic correction is considered perturbatively. The solution for the free oscillator is
\begin{align}
    q(t)= Q_i \cos \left( t\right) + P_i \sin \left( t \right) \,, \quad p(t)= P_i \cos \left( t\right) - Q_i \sin \left( t \right). \,
\end{align}
Now go to the interaction picture $(q,p)\rightarrow (Q,P)$,
\begin{align}
    Q (t) = q(t) \cos (t) - p(t) \sin (t) \,, \quad P (t) = p(t) \cos (t) + q(t) \sin (t) \,,
\end{align}
which takes us to the new Hamiltonian,
\begin{align}
    \quad \h' = \frac{\lambda}{4} \left( Q(t) \cos (t) - P(t) \sin (t) \right)^2 \,.
\end{align}
By perturbatively integrating eq. \eqref{eq:eominintpic}, we obtain the solution $(Q(t), P(t))$. Inserting this into the action, at $\mathcal{O}(\lambda)$ we get,
\begin{align}
\begin{aligned}\label{eq:anHOS-OSA}
    \s (T,0,Q_i,P_i)  &=  \frac{\lambda }{128}\left( 12 (2 P_i^3Q_i + 3P_i^4T + 2P_i^2Q_i^2T - Q_i^4T) - 32P_i^3Q_i \cos(2T) + 8P_i^3Q_i\cos(4T) \right. \\
    & \left. \quad\quad\quad\quad\quad- 8(3P_i^4 + Q_i^4)\sin(2T) + (3P_i^4 - 6P_i^2Q_i^2 - Q_i^4)\sin(4T)\right) + \mathcal{O}(\lambda^2) \,.
\end{aligned} 
\end{align}
Following the same algorithm as in the previous section we can also compute the Magnusian perturbatively and it is given to $\mathcal{O}(\lambda)$ as,
\begin{align}
\begin{aligned}
    \chi (T,0,Q_i,P_i) &= -\frac{\lambda }{128}\left( 4(3P_i^3Q_i + 5P_iQ_i^3 + 3(P_i^2 + Q_i^2)^2T - 4P_iQ_i(P_i^2 + Q_i^2))\cos(2T) \right. \\
    &  \quad\quad\quad\quad\quad + 4P_i(P_i - Q_i)Q_i(P_i + Q_i)\cos(4T)   + 8(-P_i^4 + Q_i^4)\sin(2T)  \\
    &  \quad\quad\quad\quad\quad \left. +  (P_i^4 - 6P_i^2Q_i^2 + Q_i^4)\sin(4T) \right) + \mathcal{O}(\lambda^2) \,,
\end{aligned}
\end{align}
which, again is clearly different that the on-shell action in eq. \eqref{eq:anHOS-OSA}. Computing the term relating the two to $\mathcal{O}(\lambda)$, 
\begin{align}
\begin{aligned}
    \int_0^1 p \frac{d q}{d\varepsilon}d\eps &= \frac{P_i\lambda}{32}\bigg(9P_i^2Q_i + 5Q_i^3 + 12P_i(P_i^2 + Q_i^2)T - 4Q_i(3P_i^2 + Q_i^2)\cos(2T) \\
    &\quad\quad\quad\quad- Q_i(-3P_i^2 + Q_i^2)\cos(4T) - 8P_i^3\sin(2T) + P_i(P_i^2 - 3Q_i^2)\sin(4T)\bigg) + \mathcal{O}(\lambda^2) \,,
\end{aligned}
\end{align}
which is exactly the difference between the two. We have checked the relation \eqref{eq:eik2osa} through $\mathcal{O}(\lambda^5)$ in the attached Mathematica notebook.

\subsection{Integrable Scattering} \label{sec:intsys}

In the case of integrable scattering, the Magnusian $\chi$ and the radial action $I$ coincide. Ultimately, this is because for such scattering it is possible to go to action-angle variables, in which case we can use \eqref{eq:integrablesimple} so that the on-shell action becomes a Legendre transformation of $\chi$, however it is insightful to go through the details a bit more carefully. This simple relationship is physically relevant, because the two-body problem in the conservative sector of scalar PM or PN scattering is integrable, as is scalar probe motion in a Kerr background. However, it breaks down in general for the two-body problem due to loss of integrability when spin (outside of the probe limit\footnote{Probe motion in a Kerr background to linear order in the spin of the probe remains integrable. Integrability at quadratic order in spin is maintained if and only if the probe has the same spin-induced quadrupole as a black hole~\cite{Compere:2023alp}. It is unknown if integrability can be maintained for the probe beyond quadratic order in spin for a particular selection of Wilson coefficients, but certainly already at quadratic order in spin, integrability fails for a generic body. However, it is possible that a less restricted definition of integrability (asymptotic integrability) may hold for special values of Wilson coefficients at higher orders in the probe spin, which could serve as a defining feature of idealized spinning black holes~\cite{Akpinar:2024meg,Akpinar:2025bkt,Akpinar:2025tct}.}) or radiation are included. In general, for a non-integrable system, there isn't even a meaningful definition of the radial action (that has any sense of canonical invariance) because the Hamilton-Jacobi equation is non-separable and the phase space lacks the necessary invariant tori. 

Let us study how $\chi = I$ arises for integrable scattering by looking at general motion under a spherical potential in detail. This example includes the case of conservative scattering in the center of momentum frame of two spinless bodies in the PM or PN expansions. The Hamiltonian $\h(\vec x, \vec p)$ takes the restricted form $\h(r, p)$ where $r = |\vec x|$ is the distance between the bodies and $p = |\vec p|$ is the magnitude of either body's linear momentum in the center of momentum frame. The motion lies in a plane and we use cylindrical coordinates, so $p = \sqrt{p_r^2 + \frac{p_\varphi^2}{r^2}}$. Further, we require that the Hamiltonian takes the form:
\begin{equation}
    \h(r, p) = T(p) + \frac{1}{r}U\left(\frac{1}{r}, p\right) \label{eq:scatteringham}
\end{equation}
where $T$ and $U$ are both smooth functions of their arguments. This form guarantees the particles become free no more slowly than the Coulomb interaction. Because the arguments of $T$ and $U$ are always positive as evaluated here, we define them to be even functions when evaluated at general arguments. The dynamics are integrable and give the on-shell action:
\begin{equation}
    \s(t_f, r_f, \varphi_f, t_i, r_i, \varphi_i) = R(E, L, r_f, r_i) + L(\varphi_f - \varphi_i) - E(t_f - t_i)
\end{equation}
where, by the Hamilton-Jacobi conditions:
\begin{align}
    t_f - t_i &= \frac{\d R}{\d E},     &       p_{rf} &= \frac{\d R}{\d r_f},          &       p_{\varphi f} &= L,         &       E &= \h\left(r_f, \sqrt{\left(\frac{\d R}{\d r_f}\right)^2 + \frac{L^2}{r^2_f}}\right) \nonumber\\
    \varphi_f - \varphi_i &= -\frac{\d R}{\d L},        &          p_{ri} &= -\frac{\d R}{\d r_i},      &       p_{\varphi i} &= L,     &      E &= \h\left(r_i, \sqrt{\left(\frac{\d R}{\d r_i}\right)^2 + \frac{L^2}{r^2_i}}\right). \label{eq:inthamjac}
\end{align}
The restricted form of the Hamiltonian guarantees it can be written in the form:
\begin{equation}
    \h(r, p) = T(p_r) + \frac{1}{r}U(0, p_r) + \frac{1}{r^2}V\left(\frac{1}{r}, p_r, L\right)
\end{equation}
with $V$ a smooth function of its arguments. Thus, the energy Hamilton-Jacobi condition can be inverted to give:
\begin{equation}
    \left|p_r\right| = \wp(E) + \frac{1}{r}A(E) + \frac{1}{r^2}B\left(\frac{1}{r}, E, L\right) \label{eq:hamrjac}
\end{equation}
where $\wp(E) = T^{-1}(E)$ is understood to be the positive branch of the inverse function of $T$ and $A$ and $B$ are smooth functions of their arguments. Let $\mc{r}(E, L)$ be the turning point of the trajectory, meaning the value of $r$ at which $p_r = 0$. Then:
\begin{equation}
    0 = \wp(E) + \frac{1}{\mc{r}(E, L)}A(E) + \frac{1}{\mc{r}^2(E, L)}B\left(\frac{1}{\mc{r}(E, L)}, E, L\right). 
\end{equation}
Scattering involves moving from a large $r_i$ to $\mc{r}$ then from $\mc{r}$ back up to a large $r_f$. From $r_i$ to $\mc{r}$, $p_r$ is negative while from $\mc{r}$ to $r_f$ it is positive. Integrating \eqref{eq:hamrjac} along the trajectory:
\begin{equation}
    R(E, L, r_f, r_i) = \wp(E)(r_f + r_i - 2\mc{r}) + A(E)\ln\frac{r_f r_i}{\mc{r}^2} + \int_{\mc{r}}^{r_f}\frac{B}{r^2} dr + \int_{\mc{r}}^{r_i} \frac{B}{r^2} dr. 
\end{equation}
For constants $\Lambda$ and $\mu$ both smaller than both $r_f$ and $r_i$ and larger than the turning point $\mc{r}$, and with $\vartheta$ the Heaviside step function, this can be written as:
\begin{equation}
    R(E, L, r_f, r_i) = \wp(E)(r_f + r_i - 2\Lambda) + A(E)\ln\frac{r_f r_i}{\mu^2} + R_{\Lambda,\mu}(E, L, r_f, r_i)
\end{equation}
where we have introduced the regulated radial function $R_{\Lambda,\mu}(E, L, r_f, r_i)$:
\begin{equation}
    R_{\Lambda,\mu}(E, L, r_f, r_i) = \sum_{X \in \{r_i, r_f\}} \int_{\mc{r}}^{X}\left(\wp(E) \vartheta(\Lambda-r) + \frac{A(E)}{r}\vartheta(\mu - r) + \frac{B}{r^2}\right) dr
\end{equation}
$R_{\Lambda,\mu}(E, L, r_f, r_i)$ has perfectly healthy $r_f\to\8$ and $r_i\to\8$ limits while the remaining pieces of $R(E, L, r_f, r_i)$ are $L$ independent. Therefore, by \eqref{eq:inthamjac}, the radial action $I$, defined by:
\begin{equation}
    I(E, L, \Lambda, \mu) = \lim_{\substack{r_f \to \8 \\ r_i \to \8}} R_{\Lambda,\mu}(E, L, r_f, r_i)
\end{equation}
satisfies the Hamilton-Jacobi condition:
\begin{equation}
    \varphi_f - \varphi_i = -\frac{\d I}{\d L}
\end{equation}
in the scattering case. 

We now go to the interaction picture. In order to do so, we need to appropriately regulate the Hamiltonian \eqref{eq:scatteringham} so that the $t_f\to\8$ and $t_i\to -\8$ limits will be well-behaved. In accordance with the regulated radial function $R_{\Lambda, \mu}$, we define the regulated Hamiltonian $\h_{\Lambda,\mu}$ so that:
\begin{equation}
    \h_{\Lambda,\mu}\left(r, \sqrt{p_r^2 + \frac{L^2}{r^2}}\right) = E
\end{equation}
whenever
\begin{equation}
    |p_r| = \frac{1}{r^2}B'\left(\frac{1}{r}, E, L\right)
\end{equation}
where $B'$ is the bounded function of its arguments defined by:
\begin{equation}
    B'(x, E, L) = B(x, E, L) + \frac{A(E)}{x} \vartheta\left(x - \frac{1}{\mu}\right) + \frac{\wp(E)}{x^2}\vartheta\left(x - \frac{1}{\Lambda}\right). 
\end{equation}
This essentially turns off the Coulomb potential at $r = \mu$ and turns off the free particle Hamiltonian at $r = \Lambda$. Now consider ``free-scattering'', meaning under just the time evolution generated by $T(p)$. Then, the resulting on-shell action is:
\begin{align}
    \s_0(t_f, r_f, \varphi_f, t_i, r_i, \varphi_i) =&\ \sqrt{\wp^2 r_f^2 - L^2} + \sqrt{\wp^2 r_i^2 - L^2} + L \arctan\left(\frac{L}{\sqrt{\wp^2 r_f^2 - L^2}}\right) \nonumber \\
    &\ + L\arctan\left(\frac{L}{\sqrt{\wp^2 r_i^2 - L^2}}\right)+ \left(\varphi_f - \varphi_i - \pi\right)L - E(t_f- t_i).
\end{align}
The interaction picture on-shell action is given by use of $-\s_0$ as the generating function of the canonical transformation with the transformed and untransformed variables having the same initial conditions:
\begin{equation}
    \s_I(t_f, r_{If}, \varphi_{If}, t_i, r_{Ii}, \varphi_{Ii}) = \s(t_f, r_f, \varphi_f, t_i, r_{Ii}, \varphi_{Ii}) - \s_0(t_f, r_f, \varphi_f, t_i, r_{If}, \varphi_{If}) \label{eq:intscatteract}
\end{equation}

Now we want to take the $t_f \to \8$ and $t_i \to -\8$ limits. So, we treat $t_f \sim |t_i| \sim r_i \sim r_f \sim \ell$ with $\ell$ a large parameter. Taking the $\ell\to\8$ limit of \eqref{eq:intscatteract} with everything computed using the regulated Hamiltonian gives precisely:
\begin{equation}
    \lim_{\ell\to\8} \s_I = I(E, L, \Lambda,\mu) + L(\varphi_f-\varphi_i). \label{eq:intact2I}
\end{equation}
But, by \eqref{eq:eik2osaformal}:
\begin{equation}
    \lim_{\ell\to\8} \s_I = \frac{e^{\{\chi,\bt\}}-1}{\{\chi,\bt\}}(L\{\chi, \varphi_i\}) + \chi \label{eq:intact2chi}
\end{equation}
where $\chi$ is the Magnusian corresponding to the regulated Hamiltonian. Due to the regulated Hamiltonian having no explicit $\varphi$ dependence, $\chi$ has no such dependence either, leaving it only as a function of $L, E$ (with $E$ a constant parameter on the ($\varphi, L$) phase space). Consequently, $L\{\chi, \varphi_i\} = -L\frac{\d\chi}{\d L}$ is a function of $L$ only and all of the tower of Poisson brackets trivializes. Equation \eqref{eq:eikexpdef} gives $\varphi_f - \varphi_i = -\frac{\d\chi}{\d L}$ and so matching \eqref{eq:intact2I} and \eqref{eq:intact2chi}, we find:
\begin{equation}
    \chi = I. 
\end{equation}
This same analysis (due to \eqref{eq:integrablesimple}) can be extended to more general but still integrable cases, because then \eqref{eq:intact2I} is only modified by allowing $I$ to depend on the associated additional adiabatic invariants and $L(\varphi_f - \varphi_i)$ becomes one of a sum of such terms of the form adiabatic invariant times a change in the associated angle variable. In this more general case, $\chi$ can still only depend on the adiabatic invariants, the Poisson brackets still trivialize in those variables, and the result $\chi = I$ is immediately recovered.

\section{Conclusion}\label{sec:conclusion}

In this article we have shown that the ``eikonal'' refers to two fundamentally different quantities in present literature, one which is the logarithm of an S-matrix element and the other which is a matrix element of the logarithm of the S-matrix. In the classical limit, these become the on-shell action $\s$ (eikonal phase) and the Magnusian $\chi$ (eikonal generator), respectively. We showed that in the classical limit they are precisely related by:
\begin{equation}
    \s(t_f, Q_f, t_i, Q_i) = \int_0^1\left(p \frac{d q}{d\varepsilon} + \chi(t_f, t_i, q, p)\right) d\varepsilon.
\end{equation}
We hope to have shed light on the subtle difference between the two. It is specifically the Magnusian $\chi$ which is of central usefulness in calculating observables efficiently, because it is a single scalar function which through \eqref{eq:eikexpdef} can be used to compute the impulse in any observable. While the matrix elements of the quantum S-matrix have a generally challenging classical limit, requiring care regarding $\hbar$-over-$\hbar$ pieces to ensure cancellation of superclassical effects, the eikonal operator $\hat\chi$ has a well-defined classical limit. This allows us to bring the powerful idea of the S-matrix from quantum mechanics into classical physics in the form of the classical S-matrix $e^{\{\chi,\bt\}}$. 

In sec.~\ref{sec:intsys} we showed that the radial action $I$ and the Magnusian $\chi$ coincide for planar scattering and argued that this equality holds in the generic integrable scattering case. This causes the on-shell action $\s$ and Magnusian $\chi$ to be related by a simple Legendre transform. Outside of the integrable case, such as when spin is included, that simple Legendre relationship breaks down. But, even for generic scattering, there remain relations between the two at the level of Feynman diagrams. The Magnusian can be computed using Feynman-like tree diagrams in a worldline description of massive particles using causal propagators~\cite{Kim:2024svw}, as opposed to the on-shell action which is naturally computed using Feynman diagrams through the Dyson series with Feynman propagators. The diagrams computing the two are related in that the full integrand only differs by different weightings of the propagator $i0^+$ prescriptions, which can be stated as a sum rule for the graph function $\omega(\tau)$ (which computes the weights associated to the Feynman-like graphs for the Magnusian) known as the Feynman reduction property; see eq.~(4.20) of Ref.~\cite{Kim:2024svw}. This mechanism allows the use of unitarity based methods to construct integrands for the Magnusian, at least when a worldline description is adopted~\cite{Haddad:2025cmw}. Understanding the implications of this diagrammatic relationship between the on-shell action and the Magnusian we leave as a direction for future work. 

While interest in the Magnusian has recently been motivated by scattering amplitudes applications, its existence is entirely generic to classical mechanics. In general, it acts as a resummed/time-averaged Hamiltonian over the interval of consideration, giving the correct final state in phase space derived from the true Hamiltonian, but through a time-independent flow that is a near-identity transformation of the physical trajectory. Because the Magnusian exists even outside of the context of scattering, it is a promising future direction to consider its application to bound orbits, where it would naturally become a function of orbital parameters and its derivatives would encode conjugate orbital impulses, relating the binding energy and orbital period, angular momentum and orbital precession, etc. 

\acknowledgments

The authors thank Zvi Bern, Francisco Blanco, Joon-Hwi Kim, Manfred Kraus, Sangmin Lee, Felix Lichtner, Gustav Mogull, Richard Myers, Jan Plefka, Rafael Porto, Radu Roiban, and Mao Zeng for valuable discussions. 
R.P.’s research is funded by the Deutsche Forschungsgemeinschaft (DFG, German Research Foundation), Projektnummer 417533893/GRK2575 “Rethinking Quantum Field Theory”.

\appendix 

\section{Classical Limit of Wigner Functions} \label{sec:wigner}
	
	The Weyl transform provides a clear prescription for how to write operators in a canonical ordering of $\hat Q$ and $\hat P$. Define the symmetrized operator product $\text{Sym}(\hat{\mathcal{O}}_1, ..., \hat{\mathcal{O}}_n)$ as $\frac{1}{n!}$ times the sum over all $n!$ ways to order these operators in a product. If $G(Q, P)$ is an analytic function, then series expanding it in $Q, P$ about a point $q, p$ allows the $Q, P$ integrals in \eqref{eq:weyl} to be performed. Afterward, the $k, \ell$ integrals can be performed, and doing so gives:
	\begin{equation}
		\hat G = \sum_{\substack{m=0 \\ n=0}}^{\8,\8}\frac{1}{m! n!}\frac{\partial^{m+n} G(q, p)}{\partial q^m \partial p^n}\text{Sym}(\hat Q-q,(m\text{ copies}), \hat P-p,(n\text{ copies})).	\label{eq:symmetrize}
	\end{equation}
	This is just the Taylor series for the function $G(q, p)$ formally evaluated at $(\hat Q, \hat P)$ where the prescription for how to resolve any ordering ambiguities is to fully symmetrize the operator products. 
	
	The classical limit of a quantum system is the limit in which the wavefunction for the system is sufficiently localized in both coordinate space and momentum space relative to the relevant experimental resolution to be treated as being described solely by its expected coordinates $Q_\cl =\langle \hat Q\rangle $ and expected momenta $P_\cl = \langle \hat P\rangle $. This is under the assumption we have, as previously stated, chosen our coordinates and momenta such that their expectations are classically finite.
	
    Let $M_{mn}$ be the central moments of the wavefunction of our system defined as:
    \begin{equation}
        M_{mn} = \<\psi|\sym(\hat Q-Q_\cl, (m \text{ copies}), \hat P-P_\cl, (n \text{ copies}))|\psi\>.
    \end{equation}
    In the classical limit all of these moments effectively are squeezed to 0. The finite value of $\hbar$ prevents the wavefunction from being simultaneously localized in position and momentum space, so $\hbar\to 0^+$ is necessary but not sufficient for the classical limit. After taking $\hbar\to 0^+$, we must also consider a scaling of the wavefunction so that it properly localizes. 

    Define the dilation generator $\hat D$ and the dilated wavefunction $|\psi_s\>$ by:
	\begin{equation}
		\hat D = \frac{(\hat Q-Q_\cl)(\hat P-P_\cl)+(\hat P-P_\cl)(\hat Q-Q_\cl)}{2}, \p |\psi_s\rangle  = e^{-\frac{i}{\hbar}s\hat D}|\psi\rangle.
	\end{equation}
    Let $M_{mn}(s)$ be the moments computed using $|\psi_s\>$ as the wavefunction. Then:
    \begin{equation}
        M_{mn}(s) = e^{s(m-n)} M_{mn}(0). 
    \end{equation}
    The position space spread of the wavefunction is $M_{20}(s)$, which we want to send to 0 to localize the wavefunction in the classical limit. This requires we send $s\to -\8$.
	
	Localizing the wavefunction in position space will antilocalize it in momentum space unless we first take $\hbar\to 0^+$. To implement this correctly we need to keep the expected momentum $P_\cl$ of the wavefunction fixed. We can strip off the expected momentum and leave behind a wavefunction $|\phi\>$ of expected momentum 0 by defining:
	\begin{equation}
        |\phi\> = e^{-\frac{i}{\hbar}P_\cl \hat Q} |\psi\> \implies \langle \phi|\hat P|\phi\rangle  = 0.
	\end{equation}
	The momentum operator acting on $|\phi\rangle$ contains a factor of $\hbar$ in its definition and so it is useful to define the $\hbar$-independent operator $\hat K = \frac{\hat P}{\hbar}$. Let $|\phi_s\rangle $ be the dilated $|\phi\rangle$ wavefunction and let $N_{mn}$ be the central moments of $|\phi\>$ using the now centered and $\hbar$-independent momentum operator $\hat K$:
    \begin{equation}
        N_{mn} = \<\phi|\text{Sym}(\hat Q\!-\!Q_\cl, (m \text{ copies}), \hat K, (n \text{ copies}))|\phi\rangle.
    \end{equation} 
    It is these $N_{mn}$ moments that stay fixed in the classical limit. Then, the expectation of \eqref{eq:symmetrize} becomes:
	\begin{equation}
		\langle \psi_s|\hat G|\psi_s\rangle  = \sum_{\substack{m=0 \\ n=0}}^{\infty, \8}\frac{\hbar^n e^{s(m-n)}}{m! n!}\frac{\partial^{m+n} G(Q_\cl, P_\cl)}{\partial Q^m_\cl \partial P^n_\cl}N_{mn} \label{eq:fullExp}
	\end{equation}
    We are now in a position to take $\hbar\to 0^+$ and then $s\to -\8$. Doing so and recalling the $\hbar$ expansion of the function $G$ gives:
	\begin{equation}
		G_\text{cl}(Q_\cl, P_\cl) = \lim_{s\to -\8}\lim_{\hbar\to 0^+}\langle \psi_s|\hat G|\psi_s\rangle  = G_0(Q_\cl, P_\cl)
	\end{equation}
    which is \eqref{eq:classicallimit}. The classical limit completely determines the $\hbar$ independent piece of the Wigner function of each operator (with a classically finite expectation value) to be the corresponding classical observable. But, the classical limit determines nothing about the $\hbar$ dependent pieces of an operator's Wigner function. This is the essence of the statement that there is no unique canonical quantization scheme. Different choices of scheme amount to different selections of $\hbar$ corrections. While no particular scheme is fixed by the classical limit alone, there is often more data about the quantum theory which is known than just the classical limit; symmetries of a quantum theory may constrain $\hbar$ dependent terms of Wigner functions. 

\section{Leading Classical Behavior of the Path Integral} \label{sec:quant}

    Define the function: 
    \begin{equation}
        R(t_f, Q_f, t_i, Q_i) = \lim_{\hbar\to 0^+} \hbar^\alpha e^{-\frac{i}{\hbar} \s(t_f, Q_f, t_i, Q_i)} U(t_f, Q_f, t_i, Q_i)
    \end{equation}
    for some as yet unspecified $\alpha$. We want to select $\alpha$ so that $R$ is well-defined and finitely nonzero. Let $H$ have $\hbar$ expansion:
    \begin{equation}
        H(t, Q, P) = \mathcal{H}_\text{cl}(t, Q, P) + \hbar H_1(t, Q, P) + \mathcal{O}(\hbar^2).
    \end{equation}
    Let ($Q_\cl(t), P_\cl(t)$) be the classical trajectory which makes the classical action stationary. Define $\delta Q(t)$ and $\delta P(t)$ by:
    \begin{equation}
        Q(t) = Q_\cl(t) + \sqrt{\hbar} \delta Q(t), \qquad P(t) = P_\cl(t) + \sqrt{\hbar} \delta P(t).  \label{eq:intchange}
    \end{equation}
    Then, expanding $\s_H$ in $\hbar$ yields:
    \begin{equation}
        \s_H[Q, P] = \s(t_f, Q_f, t_i, Q_i) + \hbar \twid{\s}[\delta Q, \delta P] + \hbar \int_{t_i}^{t_f} H_1(t, Q_\cl(t), P_\cl(t)) dt + \OO(\hbar^\frac{3}{2})
    \end{equation}
    where we have defined $\twid{\s}$ and $\twid{H}$ by:
    \begin{gather}
        \twid{\s} = \int_{t_0}^{t_f}(\delta P_a \dot{\delta Q}^a - \twid{H}(t, \delta Q, \delta P)) dt \\
        \twid{H} = \frac{1}{2}\frac{\partial^2 \mathcal{H}_\text{cl}(t, Q_\cl, P_\cl)}{\partial Q^a_\cl\partial Q^b_\cl} \delta Q^a \delta Q^b + \frac{\partial^2 \mathcal{H}_\text{cl}(t, Q_\cl, P_\cl)}{\partial Q^a_\cl \partial P_b^\cl} \delta Q^a \delta P_b + \frac{1}{2}\frac{\partial^2 \mathcal{H}_{\text{cl}}(t, Q_\cl, P_\cl)}{\partial P^\cl_a \partial P^\cl_b}\delta P_a \delta P_b.
    \end{gather}
    Returning these leading expansions to \eqref{eq:pathintegral} gives:
    \begin{align}
        U = \frac{1}{\hbar^\frac{d}{2}}e^{\frac{i}{\hbar} \s - i\int_{t_i}^{t_f} H_1 dt} \int_{\delta Q(t_i) = 0}^{\delta Q(t_f) = 0} e^{i\twid{\s} + \OO(\sqrt{\hbar})} \fc{D}_1 \delta Q \fc{D}_1 \delta P
    \end{align}
    where $\fc{D}_1 \delta Q \fc{D}_1 \delta P$ indicates we have taken the path integral measure of \eqref{eq:pathintegralmeasure} but in terms of the $(\delta Q, \delta P)$ variables and with $\hbar$ replaced by 1. 
    Now that $\hbar$s are fully under control, we can identify $\alpha = \frac{d}{2}$, so that:
    \begin{equation}
        R e^{i\int_{t_i}^{t_f} H_1 dt} = \int_{\delta Q(t_i) = 0}^{\delta Q(t_f) = 0} e^{i\twid{\s}} \fc{D}_1 \delta Q \fc{D}_1 \delta P \label{eq:hbar1}
    \end{equation}
    The result is now a path integral for an equivalent harmonic oscillator system. In particular, define the classical system with Hamiltonian $\widetilde{H}$ and canonical coordinates $\delta Q$ and $\delta P$. Then, define a quantum system with Hamiltonian operator so that its Wigner function is $\widetilde{H}$ and in that quantization set $\hbar = 1$. The right hand side of \eqref{eq:hbar1} is the path integral connecting $\delta Q(t_i) = 0$ and $\delta Q(t_f) = 0$ for that system. We can thus evaluate the right hand side here using standard results for the harmonic oscillator. Let $\widetilde{U}(t_f, \delta Q_f, t_i, \delta Q_i)$ be the matrix elements of the time evolution operator of this system. For any Hamiltonian which is a quadratic polynomial of the coordinates and momenta, even with time dependent coefficients, the matrix elements of the time evolution operator can be explicitly computed using standard quantum mechanics to be exactly:
    \begin{equation}
        \widetilde{U}(t_f, \delta Q_f, t_i, \delta Q_i) = \left(\frac{-1}{2\pi i}\right)^{\frac{d}{2}}\sqrt{\text{det} \frac{\partial^2 \widetilde{\s}}{\partial \delta Q_f \partial \delta Q_i}} e^{i \widetilde{\s}}
    \end{equation}
    where $\widetilde{\s}$ is the classical on-shell action of that system. We are interested in this at $\delta Q_i = \delta Q_f = 0$. Under these boundary conditions, $\widetilde{\s} = 0$ and the matrix appearing in the determinant is:
    \begin{equation}
        \frac{\partial^2 \widetilde{\s}}{\partial \delta Q_f^a \partial \delta Q_i^a} = \frac{1}{2}\left(\frac{\partial \delta P_{fa}}{\partial \delta Q^b_i} - \frac{\partial \delta P_{ib}}{\partial \delta Q^a_f}\right). 
    \end{equation}
    By returning to \eqref{eq:intchange} and considering varying the boundary conditions of the path, it immediately follows (using the Hamilton-Jacobi conditions) that:
    \begin{equation}
        \frac{\partial \delta P_{fa}}{\partial \delta Q^b_i} = \frac{\partial P_{fa}}{\partial Q_{i}^b} = \frac{\partial^2 \s}{\partial Q_{f}^a Q_{i}^b}, \qquad \frac{\partial \delta P_{ib}}{\partial \delta Q^b_f} = \frac{\partial P_{ib}}{\partial Q_{f}^a} = -\frac{\partial^2 \s}{\partial Q_{f}^a Q_{i}^b}
    \end{equation}
    and thus we ultimately have that the proper statement of the leading classical behavior of the matrix elements of the time evolution operator of a generic quantum system is \eqref{eq:leadingclassical}.

\section{Example: Rabi Oscillations} \label{sec:su2}

\begin{figure}[t!]
\centering
        \includegraphics[totalheight=8cm]{
        %example_affinegroup.png
        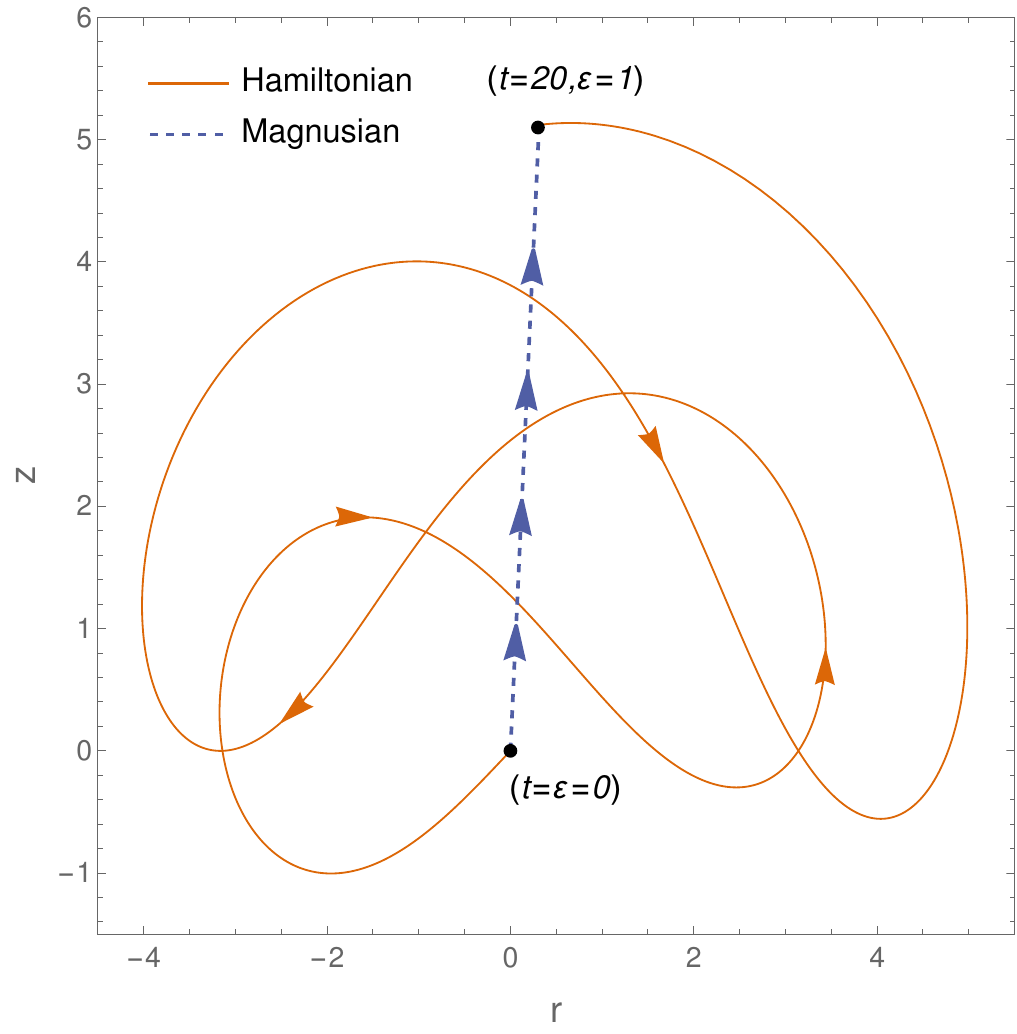
        }
    \caption{Example of group trajectories for $SU(2)$ transformations due to evolution using the physical Hamiltonian (shown in solid orange) and Magnusian (shown in dashed blue). This is computed for the case of $\Omega=\sqrt{\pi/2}$ and $\omega=1$.
    }
    \label{fig:su2group}
\end{figure}

Another good example for visualizing the Magnusian path is given by motion on $SU(2)$. Consider a complex spinor $|\psi(t)\>$. It can be given arbitrary motion through the Bloch sphere via the Euler-Lagrange equations from a Lagrangian of the form:
\begin{equation}
    \l = i\<\psi|\dot\psi\> - \<\psi|\hat H(t)|\psi\>
\end{equation}
where the Hamiltonian operator $\hat H(t)$ is an arbitrary element of the Lie Algebra $\mf{su}(2)$:
\begin{equation}
    \hat H(t) = H^a(t) \frac{\sigma_a}{2}. 
\end{equation}
A case of particular physical interest is resonant Rabi flopping, in which we take $H^a(t)$, for some constants $\omega, \Omega$, to be given by:
\begin{equation}
    H^a(t) = (\Omega \cos\omega t, \Omega \sin\omega t, \omega)^a.
\end{equation}
The resulting equations of motion are easily exactly solved, and at all times the state $|\psi(t)\>$ is related to the initial conditions by an element $\hat U(t)$ of $SU(2)$:
\begin{equation}
    |\psi(t)\> = \hat U(t) |\psi(0)\>.
\end{equation}
A generic $SU(2)$ group element is parameterized by 3 coordinates $x^a = (x,y,z)^a$ so that:
\begin{equation}
    \hat U = e^{\frac{i}{2} x^a\sigma_a}.
\end{equation}
The solution of the equations of motion thus defines a trajectory $x^a(t)$ through the group. Let $\theta(t)$ be given by:
\begin{equation}
    \theta(t) = \sqrt{x^2(t) + y^2(t) + z^2(t)}
\end{equation}
The exact group trajectory is:
\begin{equation}
    x(t) = -\frac{\theta\sin\frac{\Omega t}{2} \cos\frac{\omega t}{2}}{\sin\frac{\theta}{2}}, \p y(t) = -\frac{\theta\sin\frac{\Omega t}{2} \sin\frac{\omega t}{2}}{\sin\frac{\theta}{2}}, \p z(t) = -\frac{\theta\cos\frac{\Omega t}{2}\sin\frac{\omega t}{2}}{\sin\frac{\theta}{2}}. 
\end{equation}
$SU(2)$ is most properly visualized as the 3-sphere, but can be made easier to imagine using cylindrical-like-coordinates $(r(t), \varphi(t), z(t))$ with $x = |r|\cos\varphi$ and $y = |r|\sin\varphi$. In these coordinates, $\varphi = \frac{\omega t}{2}$ and the sign of $r$ is chosen so that $r(t)$ remains continuous and differentiable. The trajectories of evolution are shown in figure \ref{fig:su2group}, where we can see the simple sense in which the Magnusian path is geodesic.

%\bibliography{ref}{}
%\setlength{\bibsep}{0pt plus 0.1ex}
%\bibliographystyle{JHEP}

\providecommand{\href}[2]{#2}\begingroup\raggedright\endgroup

\end{document}